\documentclass[]{svjour3}
\usepackage{txfonts}
\usepackage{graphicx}
\usepackage{amssymb}
\usepackage[below]{placeins}
\usepackage{natbib}
\bibpunct{(}{)}{;}{a}{}{,}
\def \arcmin{$^{\prime}$}

\def \xmm{{\sl{XMM-Newton}}}
\def \chandra{{\sl{Chandra}}}
\def \asca{{\sl{ASCA}}}
\def \beppo{{\sl{BeppoSAX}}}

\def \xeus{{\sl{XEUS}}}
\def \leftp{{\sl{Left panel: }}}
\def \rightp{{\sl{Right panel: }}}

\def \sncc{SN$_{\mathrm{CC}}$}
\newcommand{\ion}[2]{#1\,{\sc{#2}}}

\makeatletter
\renewcommand*{\@biblabel}[1]{\hfill}
\makeatother

\journalname{SSRv}
\begin{document}

\title{Observations of metals in the intra-cluster medium}

\author{N.~Werner \and
        F.~Durret \and 
        T.~Ohashi \and
        S.~Schindler \and
	R.P.C.~Wiersma
}
\authorrunning{N. Werner et al.}

\institute{N. Werner \at 
              SRON Netherlands Institute for Space Research, Sorbonnelaan 2, NL - 3584 CA Utrecht, the Netherlands \\
             Max-Planck-Institute f{\"u}r Astrophysik, Karl-Schwarzschild-Strasse 1, 85749, Garching, Germany \\
              \email{n.werner@sron.nl}
         \and
	    F.~Durret \at
           Institut d'Atrophysique de Paris, CNRS, UMR 7095, Universit\'e Pierre et Marie Curie, 98bis Bd Arago, F-75014 Paris, France
	 \and
	 T. Ohashi \at
          Institute of Space and Astronautical Science, JAXA, 3-1-1 Yoshinodai, Sagamihara, Kanagawa 229-8510, Japan
	 \and
	 S. Schindler \at
	  Institut f{\"u}r Astro- und Teilchenphysik, Universit{\"a}t Innsbruck, Technikerstr. 25, A-6020 Innsbruck, Austria
	 \and
	 R.P.C.~Wiersma \at
	  Leiden Observatory, Leiden University, P.O. Box 9513, 
	  2300 RA Leiden, The Netherlands
             }

\date{Received: 1 October 2007 ; Accepted: 29 November 2007 }

\maketitle

\begin{abstract}

Because of their deep gravitational potential wells, clusters of galaxies retain all the metals produced by the stellar populations of the member galaxies. Most of these metals reside in the hot plasma which dominates the baryon content of clusters. This makes them excellent laboratories for the study of the nucleosynthesis and chemical enrichment history of the Universe. Here we review the history, current possibilities and limitations of the abundance studies, and the present observational status of X-ray measurements of the chemical composition of the intra-cluster medium.  
We summarise the latest progress in using the abundance patterns in clusters to put constraints on theoretical models of supernovae and we show how cluster abundances provide new insights into the star-formation history of the Universe. 
\end{abstract}

\section{Introduction}
\label{Introduction} 

Clusters of galaxies are excellent astrophysical laboratories, which allow us to
study the chemical enrichment history of the Universe. They have the deepest
known gravitational potential wells which keep the metals produced in the
stellar populations of the member galaxies within the clusters. About 70--90~\%
of the baryonic mass content of clusters of galaxies is in the form of hot
($10^7-10^8$~K) X-ray emitting gas \citep{ettori1999}. In this hot intra-cluster
medium (ICM) the dominant fraction of cluster metals resides. To the extent
that the stellar populations where the cluster metals were synthesised can be
considered representative, the metal abundances in the ICM provide constraints
on nucleosynthesis and on the star formation history of the Universe. 

It is remarkable that all the abundant elements, that were synthesised in  stars
after the primordial nucleosynthesis, have the energies of their  K- and L-shell
transitions in the spectral band accessible to modern X-ray telescopes.  Most of
the observed emission lines in the ICM arise from the well understood  hydrogen
and helium like ions and their equivalent widths can be, under the reasonable
assumption of collisional equilibrium, directly converted into the elemental
abundance of the corresponding element. Complications arising from optical depth
effects, depletion into dust grains, extinction, non-equilibrium ionisation are
minimal or absent. Therefore, abundance determinations of the hot ICM are more
robust than those of stellar systems, \ion{H}{ii} regions or planetary nebulae. This
relatively uncomplicated physical environment makes the ICM an attractive tool
for studies of the chemical enrichment (for the theoretical progress in metal
enrichment processes see \citealt{schindler2008} - Chapter~17, this volume).

\section{Sources of metals}
\label{supernovasec}

Most of the metals from O up to the Fe-group are produced by supernovae.  The
supernovae can be roughly divided into two groups: Type Ia supernovae (SN~Ia)
and core collapse supernovae (\sncc). 

\begin{figure}  
\begin{center}
\includegraphics[height=0.9\textwidth,clip=t,angle=270]{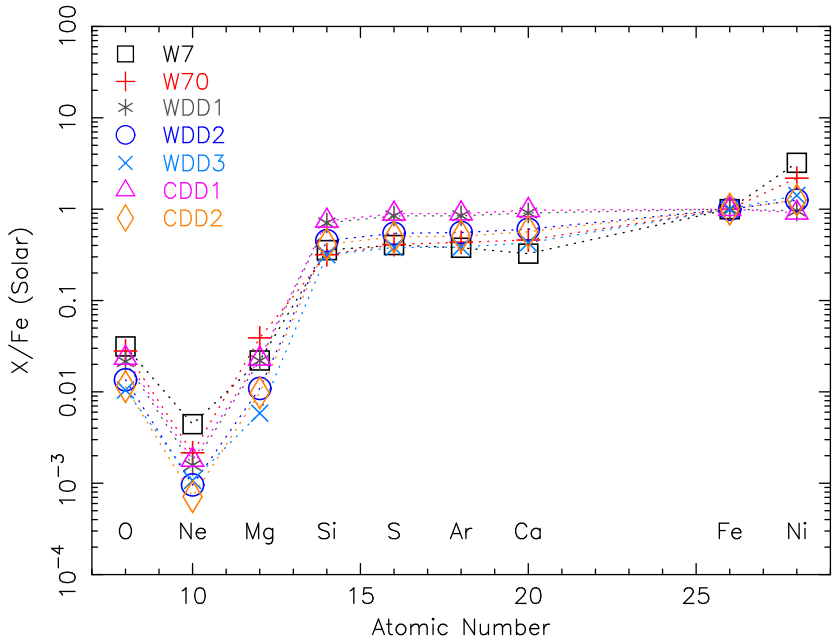}
\end{center}
\caption{Yields of elements relative to Fe in the Solar units of
\citet{grevesse1998} for different SN~Ia models from \citet{iwamoto1999}.}
\label{SNIa}
\end{figure}

\begin{table}[!htbp]
\caption{Yields of elements relative to Fe in the Solar units of
\citet{grevesse1998} for different SN~Ia models from \citet{iwamoto1999}.}
\label{tab:sn1a}
\begin{tabular}{l*7{r@{.}l}}
\hline\noalign{\smallskip}
Element    & \multicolumn{2}{c}{W7}& \multicolumn{2}{c}{W70}& \multicolumn{2}{c}{WDD1} &
\multicolumn{2}{c}{WDD2}& \multicolumn{2}{c}{WDD3}& \multicolumn{2}{c}{CDD1}& \multicolumn{2}{c}{CDD2}\\
\hline\noalign{\smallskip}
O &    0&0314     & 0&0280   	&  0&0214      &      0&0136           & 0&0105        &  0&0235       & 0&0114   \\
Ne&    0&00443    & 0&00215     &  0&00157     &      0&000960         & 0&00107       &  0&00180      & 0&000714   \\
Mg&    0&0221     & 0&0390   	&  0&0219      &      0&0109           & 0&00584       &  0&0231       & 0&00976   \\
Si&    0&356      & 0&317   	&  0&706       &      0&453            & 0&316         &  0&746        & 0&413   \\
S &    0&405      & 0&410   	&  0&843       &      0&544            & 0&380         &  0&890        & 0&496   \\
Ar&    0&378      & 0&433   	&  0&846       &      0&555            & 0&384         &  0&896        & 0&506   \\
Ca&    0&326      & 0&460   	&  0&908       &      0&604            & 0&427         &  0&972        & 0&566   \\
Fe&$\equiv$1&0 &$\equiv$1&0 &$\equiv$1&0 &$\equiv$1&0 &$\equiv$1&0 &$\equiv$1&0 &$\equiv$1&0 \\
Ni&    3&22       & 2&18   	&  0&957       &      1&25   	       & 1&41          &  0&914        & 1&26   \\
\noalign{\smallskip}\hline
\end{tabular}
\end{table}

SN~Ia are most likely thermonuclear explosions of accreting white dwarfs. When
the white dwarf reaches the Chandrasekhar limit, carbon ignition in the central
region leads to a thermonuclear runaway. A flame front then propagates through
the star at a subsonic speed as a deflagration wave. In the delayed-detonation
models, the deflagration wave is assumed to be transformed into a detonation at
a specific density.   SN~Ia produce a large amount of Fe, Ni, and Si-group
elements (Si, S, Ar, and Ca). Contrary to \sncc, they produce only very small
amounts of O, Ne, and Mg. In Fig.~\ref{SNIa} and Table~\ref{tab:sn1a} we show the theoretically
calculated yields for different SN~Ia models \citep{iwamoto1999}. We show the
yields of elements relative to the yield of Fe in the Solar units of
\citet{grevesse1998}.  Convective deflagration models are represented by the W7
and W70 models. The WDD1, WDD2, WDD3, CDD1, and CDD2 models refer to
delayed-detonation explosion scenarios and the last digit indicates the
deflagration to detonation density in units of $10^7$~g~cm$^{-3}$. The ``C'' and
``W'' refer to two different central densities ($1.37\times 10^9$ and
$2.12\times 10^9$~g\,cm$^{-3}$, respectively) in the model at the onset of the
thermonuclear runaway. The relative yield of the Si-group elements can be a good
indicator of the incompleteness of the Si-group burning into Fe-group elements.
Delayed detonation models have typically higher yields of Si-group elements
relative to Fe than the deflagration models. The Ni/Fe abundance ratio in the
SN~Ia ejecta is an indicator of neutron-rich isotope production, which depends
on the electron capture efficiency in the core of the exploding white dwarf. The
deflagration models produce larger Ni/Fe ratios than the delayed detonation
models.  

\begin{figure}
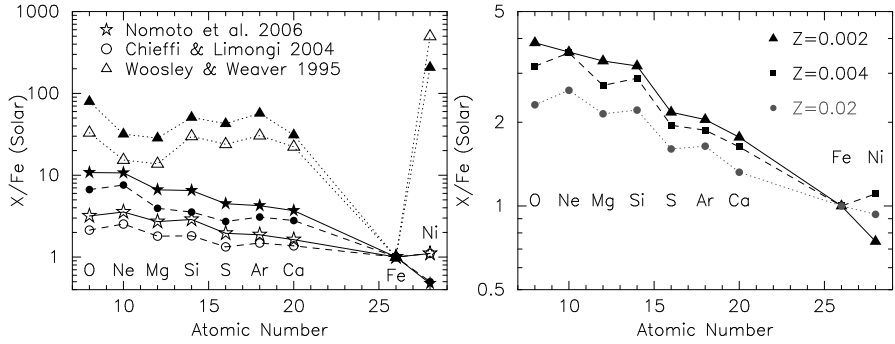
   
\includegraphics[height=0.48\textwidth,angle=270,clip=]{fig2a.ps}
\includegraphics[height=0.48\textwidth,angle=270,clip=]{fig2b.ps}
\caption{\leftp\ Yields of elements relative to Fe in the Solar units of \citet{grevesse1998} for different \sncc\ models
with a progenitor metallicity of $Z=0.004$ integrated over Salpeter and top-heavy initial mass functions between 10 and 50 M$_{\odot}$. 
\rightp\ \sncc\ yields for different progenitor metallicities \citep[$Z=0.02$ refers to Solar metallicity; ][]{nomoto2006} integrated over the Salpeter IMF. }
\label{SNcc}
\end{figure}

In the left panel of Fig.~\ref{SNcc} and Table~\ref{tab:sncc} we show the theoretical yields for three
\sncc\ models integrated over Salpeter and top-heavy initial mass functions
(IMFs) between 10 and 50 M$_{\odot}$ \citep{woosley1995,chieffi2004,nomoto2006}.
See also \citealt{borgani2008} - Chapter 18, this volume for more details on
stellar population models. In the right panel of Fig.~\ref{SNcc} we show the
\sncc\ yields for the Salpeter IMF for different progenitor metallicities
\citep[where $Z=0.02$ refers to Solar metallicity;][]{nomoto2006}. The Fe yield
of the model by \citet{woosley1995} is about an order of magnitude lower than
that in the models by \citet{chieffi2004} and \citet{nomoto2006}. The plots also
show that abundance ratios of O/Mg and Ne/Mg can be used to discriminate between
different IMFs. The \sncc\ yields for progenitors of different metallicities
differ mainly in O/Ne/Mg ratio. Because the progenitor metallicities and perhaps
also the IMF evolve, the applicability of comparing the observed data with a
single average set of yields may be limited. However, the abundances of these
elements in the ICM may still be used to infer to what extent was the
{\sl{dominant fraction}} of the \sncc\ progenitors preenriched, and what was its
IMF. 

We note, that while the abundances in Fig.~\ref{SNIa} and \ref{SNcc} are shown
with respect to the Solar values of \citet{grevesse1998}, in literature and also
in several figures of this review the abundances are often shown with respect to
the outdated Solar abundances by \citet{anders1989} or with respect to newer
sets of Solar and proto-solar abundances by \citet{lodders2003}. The more recent
Solar abundance determinations of O, Ne, and Fe by \citet{lodders2003} are 
$\sim$30~\% lower than those given by \citet{anders1989}. The Solar
abundances of O and Ne reported by \citet{grevesse1998} are higher and the
abundance of Fe is slightly lower than those reported by \citet{lodders2003}. A comparison of different Solar abundance values is shown in Table~\ref{tab:Solar}.

\begin{table}[!htbp]
\caption{Yields of elements relative to Fe in the Solar units of
\citet{grevesse1998} for different \sncc\ models 
with a progenitor metallicity of $Z=0.004$ integrated over Salpeter and 
top-heavy initial mass functions between 10 and 50 M$_{\odot}$.}
\label{tab:sncc}
\begin{center}
\begin{tabular}{l*6{r@{.}l}}
\hline\noalign{\smallskip}
Element    & \multicolumn{4}{c}{Nomoto et al. (2006)}&
\multicolumn{4}{c}{Chieffi \& Limongi (2004)}& \multicolumn{4}{c}{Woosley
\& Weaver (1995)} \\
           & \multicolumn{2}{c}{Salpeter}& \multicolumn{2}{c}{top-heavy}& \multicolumn{2}{c}{Salpeter} &
\multicolumn{2}{c}{top-heavy}& \multicolumn{2}{c}{Salpeter}& \multicolumn{2}{c}{top-heavy}\\
\hline\noalign{\smallskip}
O&           3&18    &	      	     10&8   &	  2&13   & 	6&67   &	33&0	&	79&6 \\
Ne&          3&55    &		     10&7   &	  2&53   & 	7&59   &	15&4	&	31&8 \\
Mg&          2&71    &		     6&63   &	  1&80   & 	3&94   &	13&8	&	28&5 \\ 
Si&          2&88    &		     6&50   &	  1&82   & 	3&55   &	29&9	&	50&9 \\ 
S&          1&95    &		     4&49   &	  1&33   & 	2&72   &	23&9	&	42&7 \\
Ar&          1&87    &		     4&24   &	  1&49   & 	3&08   &	30&5	&	57&1 \\
Ca&          1&63    &		     3&70   &	  1&37   & 	2&80   &	22&2	&	31&1 \\
Fe&$\equiv$1&0 &$\equiv$1&0 &$\equiv$1&0 &$\equiv$1&0   &$\equiv$1&0 &$\equiv$1&0 \\        
Ni&          1&10    &		    0&475  &	  1&11     &   0&501  &	        497&   	&	208& \\
\noalign{\smallskip}\hline
\end{tabular}
\end{center}
\end{table}

\begin{table}[!htbp]
\caption{Comparison of Solar abundances by \citet[AG;][]{anders1989},
\citet[GS;][]{grevesse1998}, and of the prot-solar abundances by
\citet{lodders2003} on a logarithmic scale with H$\equiv$12. These abundances are used at several places of this volume. }
\label{tab:Solar}
\begin{center}
\begin{tabular}{lccc}
\hline\noalign{\smallskip}
element& AG  & GS &Lodders\\
\hline
H   & 12.00&12.00&12.00\\
He  &10.99&10.93&10.98\\
C   &8.56&8.52&8.46\\
N   &8.05&7.92&7.90\\
O   &8.93&8.83&8.76\\
Ne  &8.09&8.08&7.95\\
Mg  &7.58&7.58&7.62\\
Si  &7.55&7.55&7.61\\
S   &7.21&7.33&7.26\\
Ar  &6.56&6.40&6.62\\
Ca  &6.36&6.36&6.41\\
Fe  &7.67&7.50&7.54\\
Ni  &6.25&6.25&6.29 \\
\noalign{\smallskip}\hline
\end{tabular}
\end{center}
\end{table}

While elements from O up to the Fe-group are produced by supernovae, the main
sources of carbon and nitrogen are still being debated. Both elements are
believed to originate from a wide range of sources including winds of
short-lived massive metal rich stars, longer-lived low- and intermediate-mass
stars, and also an early generation of massive stars
\citep[e.g.][]{gustafsson1999,chiappini2003,meynet2002}.  

In the Galaxy, \citet{shi2002} found that C is enriched by winds of metal-rich
massive stars at the beginning of the Galactic disk evolution, while at a later
stage it is produced mainly by low-mass stars.  \citet{bensby2006} found that
the C enrichment in the Galaxy is happening on a time scale very similar to that
of Fe. They conclude that while in the early Universe the main C contributors
are massive stars, C is later produced mainly by asymptotic giant branch (AGB)
stars.

Nitrogen is produced during hydrogen burning via the CNO and CN cycles as both a
primary and secondary element. In primary nucleosynthesis its production is
independent of the initial metallicity of the star. Primary production of N
happens during hydrogen shell burning in intermediate mass stars of
$\sim 4 - 8$~M$_{\odot}$ \citep{matteucci1985}. Stellar models that include
the effects of rotation indicate that massive stars between 9 and 20 M$_{\odot}$
may also produce primary N \citep{maeder2000}. In secondary production, which is
common to stars of all masses, N is synthesised from C and O and its abundance
is therefore proportional to the initial metallicity of the star. The primary
versus secondary production of N can be studied by investigating the N/O ratio
as a function of O/H ratio. In the case of primary N production the N/O ratio
will be constant. For the secondary production we will observe a linear
correlation between the logarithms of N/O and O/H. The combination of primary
and secondary nucleosynthesis will produce a non-linear relation. The ICM
abundance of C and N was measured only in a few bright nearby galaxy clusters
and elliptical galaxies \citep{peterson2003,werner2006b}. However, future
instruments (see Sect.~\ref{future}) will allow to measure the abundances of
these elements in many clusters. This will be important for a better
understanding of the stellar nucleosynthesis.

\section{Abundance studies before \xmm\ and \chandra}
\label{historysec}

\begin{figure}    
\begin{center}
\includegraphics[width=6.0cm,clip=t,angle=0]{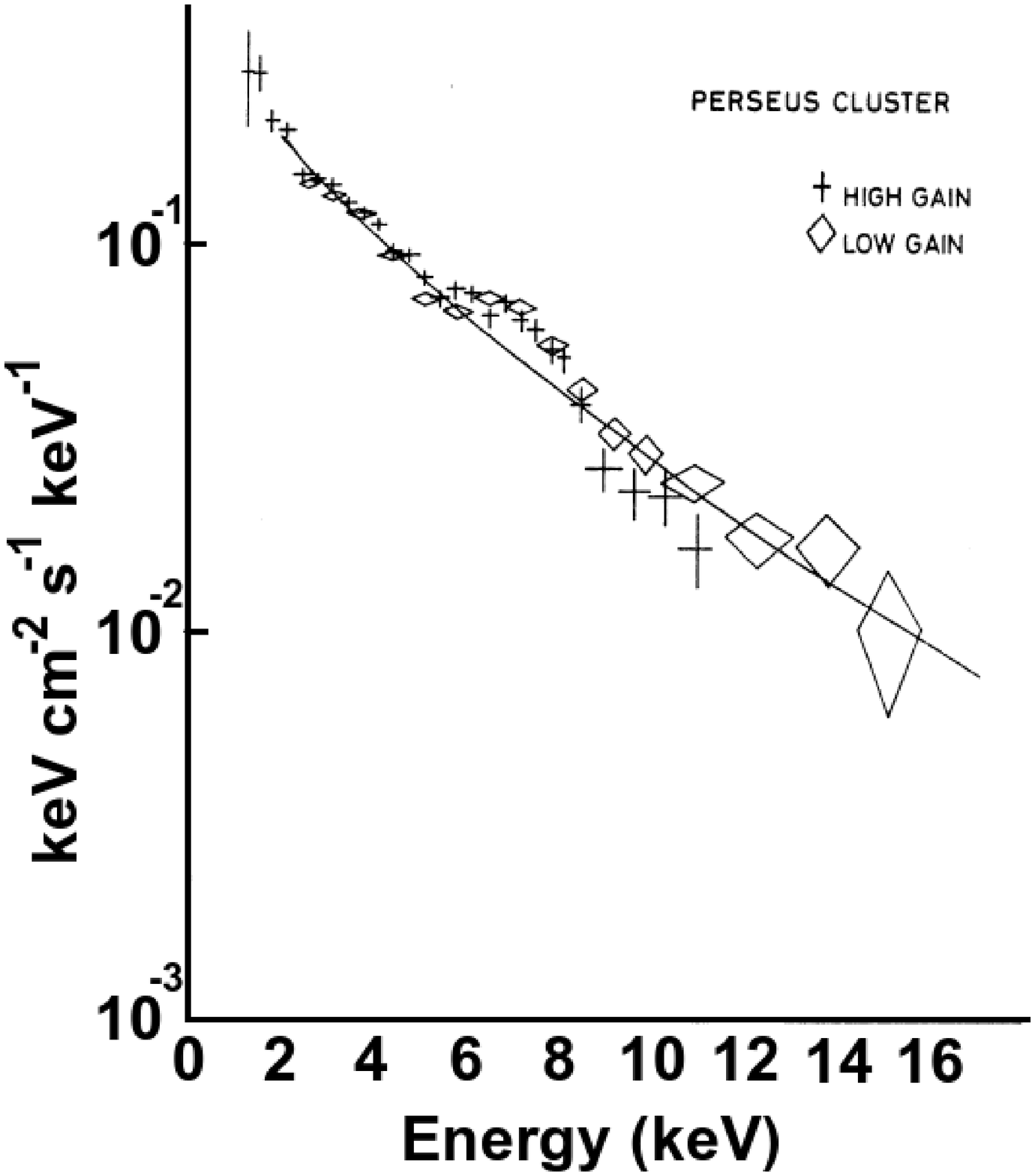}
\includegraphics[width=5.7cm,clip=t,angle=0]{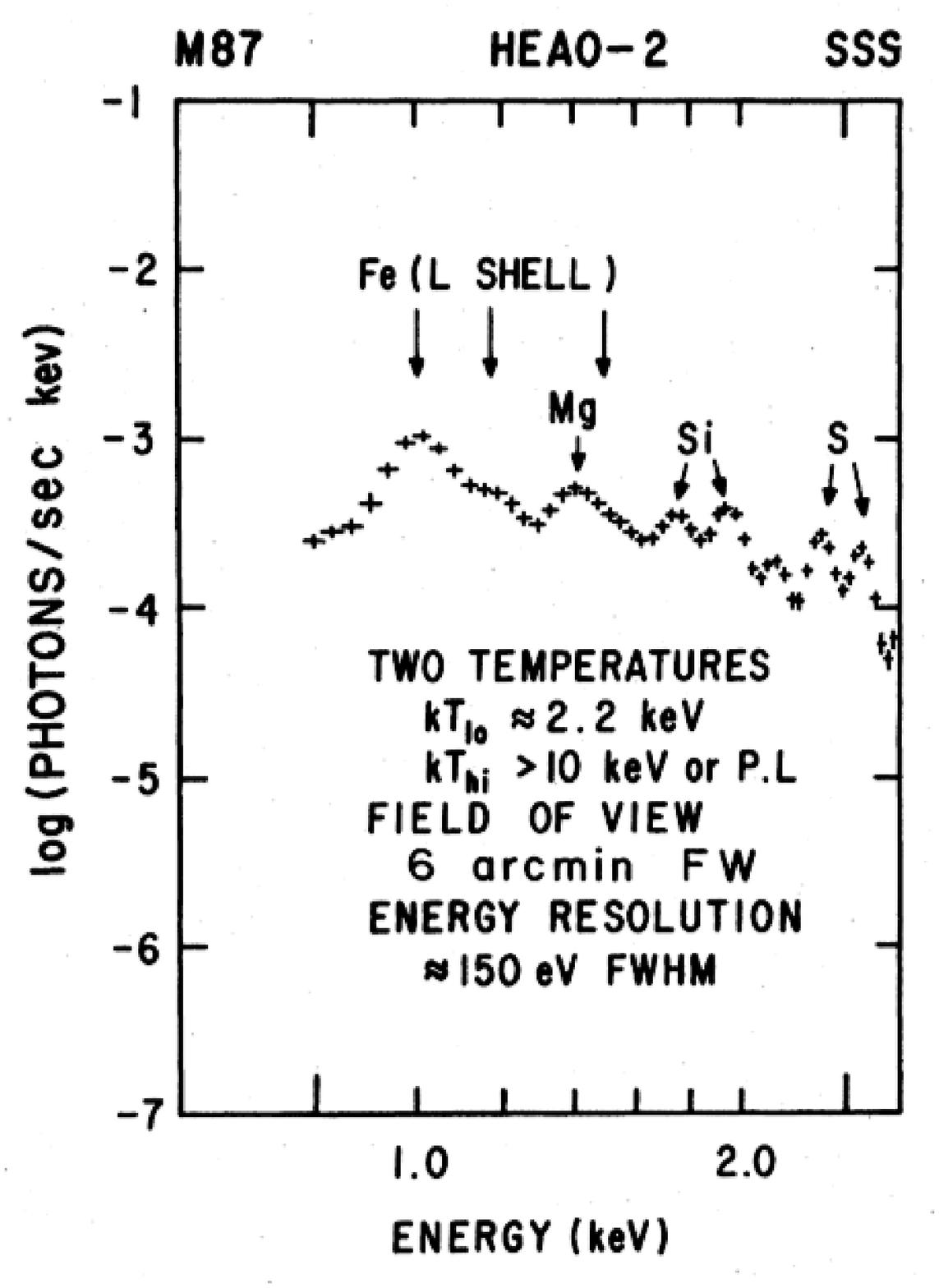}
\caption{{\emph{Left panel:}} Spectrum of the Perseus cluster obtained by {\sl{Ariel~5}}. The first cluster spectrum with observed Fe-K line emission (hump at $\sim$7~keV). From \citet{mitchell1976}. 
{\emph{Right panel:}} Line emission observed in M~87 with the {\sl{Einstein}} satellite. From \citet{sarazin1988}, based on the work published by \citet{lea1982}.}
\label{early}
\end{center}  
\end{figure}

The discovery of the Fe-K line emission in the spectrum of the Perseus cluster
by the {\sl{Ariel V}} satellite \citep[see the left panel of 
Fig.~\ref{early},][]{mitchell1976} and in Coma and Virgo by {\sl{OSO-8}}
\citep{serlemitsos1977} confirmed that the X-ray emission of galaxy clusters is
predominantly thermal radiation from hot intra-cluster gas rather than inverse
Compton radiation. These observations showed that the hot plasma in clusters of
galaxies contains a significant portion of processed gas, which was ejected from
stars in the cluster galaxies. Subsequent spectroscopic analysis of cluster
samples observed with {\sl{OSO-8}} and {\sl{HEAO-1\,A2}} satellites revealed
that the ICM has an Fe abundance of about one-third to one-half of the Solar
value \citep{mushotzky1978,mushotzky1984}. Combining spectra obtained by
{\sl{Einstein}} and {\sl{Ginga}}, \citet{white1994b} found an indication of
a centrally enhanced metallicity in four cooling flow clusters. Unfortunately,
the spectrometers on the {\sl{Einstein}} observatory were not sensitive to the
Fe-K line emission, because the mirror on the satellite was not sensitive to
photon energies above $\sim$4~keV. However, the Solid-State Spectrometer (SSS)
and the high-resolution Focal Plane Crystal Spectrometer (FPCS) on
{\sl{Einstein}} \citep{giacconi1979} allowed to detect emission lines at low
energies. Using the SSS the K lines from Mg, Si, and S and the L lines from
Fe were detected in the spectrum of M~87 (see the right panel of
Fig.~\ref{early}), Perseus, A~496, and A~576
\citep{lea1982,mushotzky1981,nulsen1982,mushotzky1984}. Using the FPCS spectra
of M~87 the \ion{O}{viii} K$\alpha$ line was detected \citep{canizares1979},
implying an O/Fe ratio of 3--5, and the relative strength of various Fe-L line
blends showed that the gas cannot be at a single temperature. 

However, until the launch of {\sl{ASCA}} in 1993, Fe was the only element for
which the abundance was accurately measured in a large number of clusters.
{\sl{ASCA}} allowed to detect the emission features from O, Ne, Mg, Si, S, Ar,
Ca, Fe, and Ni in the spectra of a number of clusters. Furthermore, it allowed
to accurately determine the Fe abundances out to redshift $z\approx0.5$.
{\sl{ASCA}} data revealed a lack of evolution in the Fe abundance out to
redshift $z\sim0.4$ \citep{mushotzky1997,rizza1998} and no evidence for a
decrease at higher redshifts \citep{donahue1999}. {\sl{ASCA}} data were the
first to show that in cooling core clusters the metallicity increases toward the
centre \citep[][]{fukazawa1994}.

\begin{figure}    
\begin{center}
\begin{minipage}{0.45\textwidth}
\hspace{-0.5cm}
\includegraphics[width=5.7cm,clip=t,angle=0]{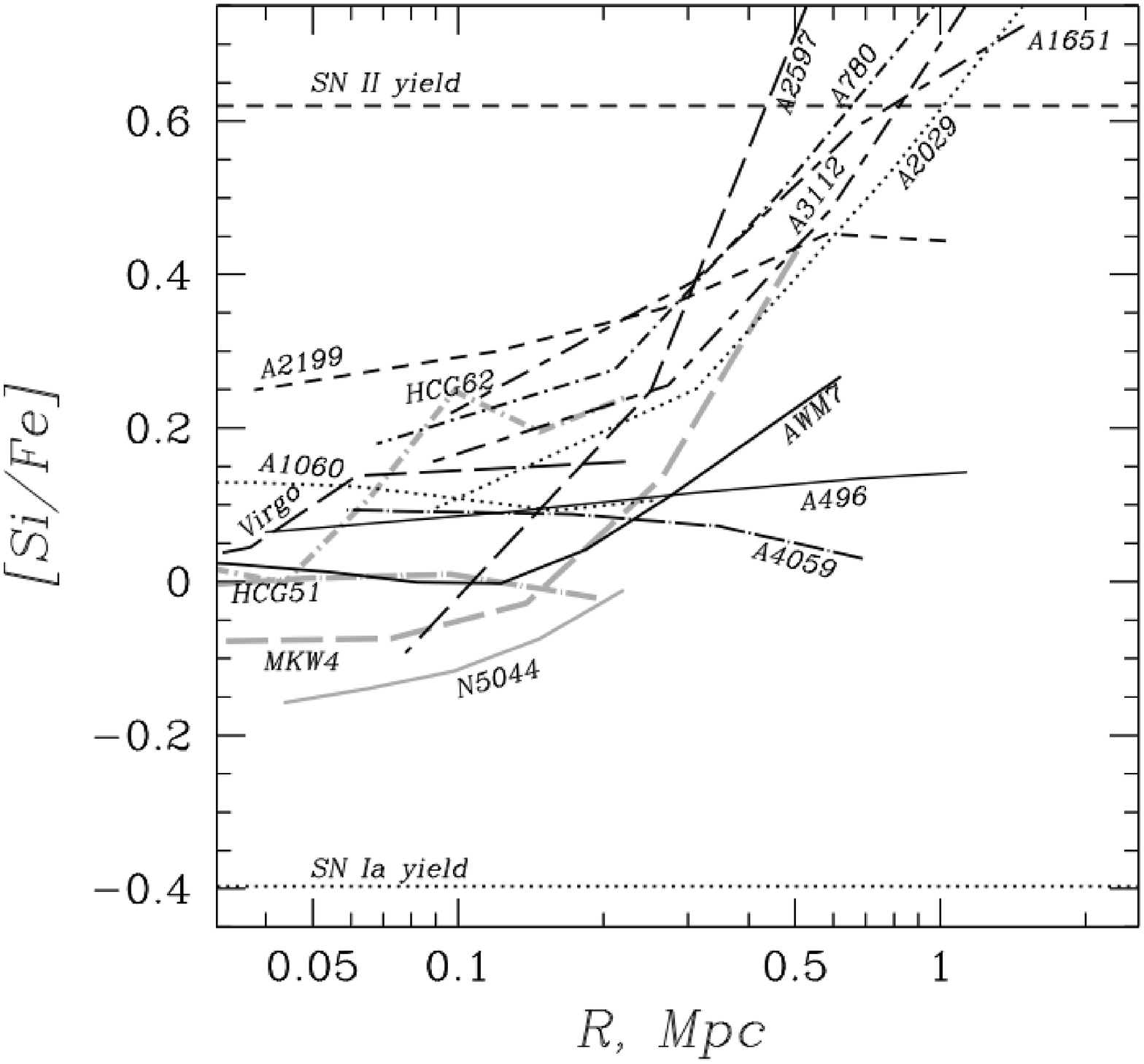}
\end{minipage}
\hspace{0.cm}
\begin{minipage}{0.45\textwidth}
\includegraphics[width=4.5cm,clip=t,angle=270]{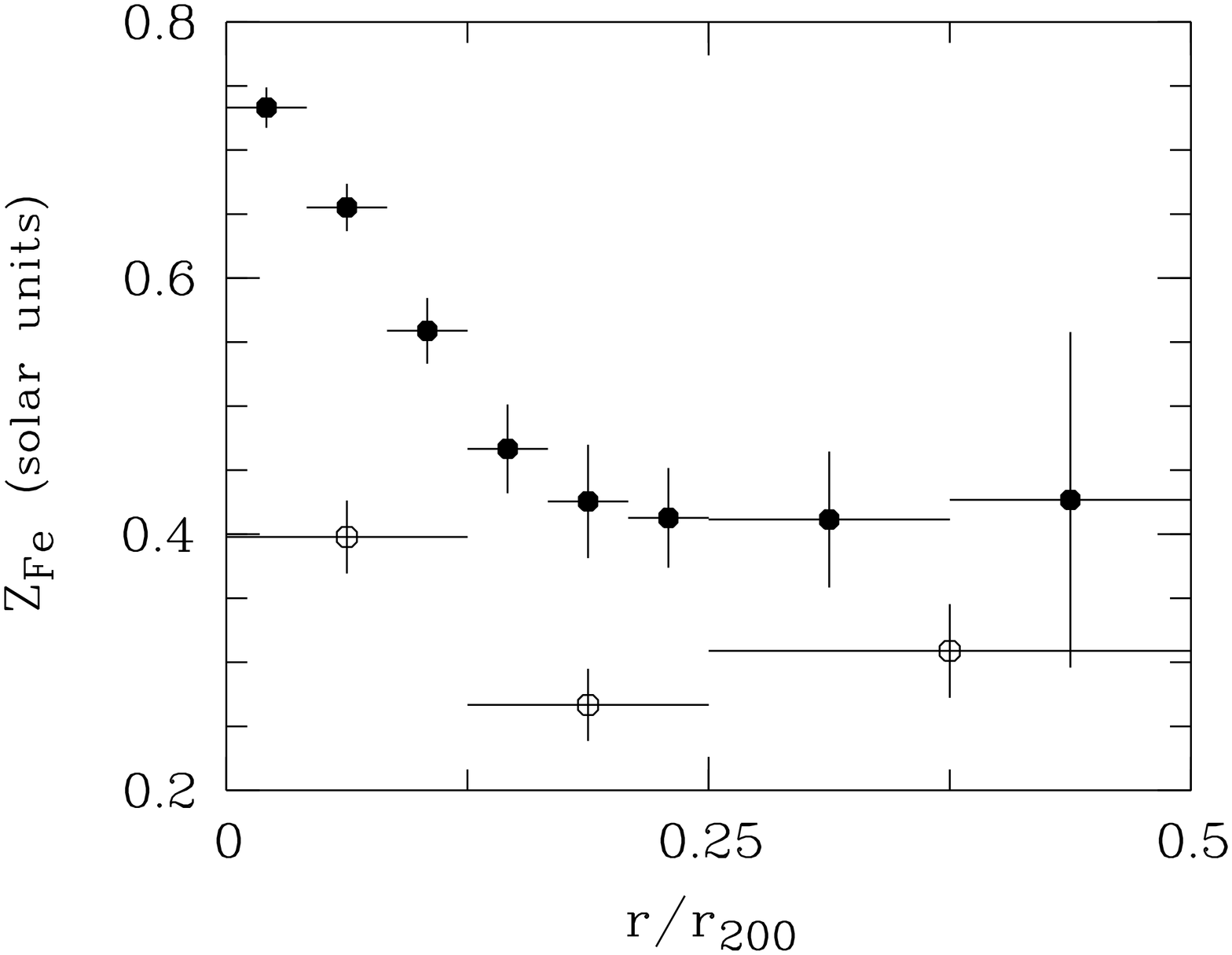}
\end{minipage}
\caption{{\emph{Left panel:}} The radial distribution of the logarithm of the
Si/Fe ratio expressed in Solar units (scale of \citealt{anders1989}) for a sample of clusters observed with
{\sl{ASCA}} \citep[from][]{finoguenov2000}.  {\emph{Right panel: }}The average
radial distribution of the Fe abundance in Solar units (scale of \citealt{grevesse1998}) for a sample of cooling core (filled
circles) and non-cooling core (empty circles) clusters observed with
{\sl{BeppoSAX}} \citep[from][]{degrandi2004}.  }
\label{radialfig}
\end{center}  
\end{figure}

By analysing data of four clusters of galaxies, \citet{mushotzky1996} found that
the relative abundances of O, Ne, Si, S, and Fe are consistent with an origin in
\sncc. They suggest that \sncc\ provided a significant fraction of metals in the
ICM. \citet{fukazawa1998} showed that the Si abundance and the Si/Fe ratio
increases from the poorer to the richer clusters, suggesting that the relative
contribution of \sncc\ increases toward richer clusters. They propose the
possibility that a considerable fraction of \sncc\ products escaped the poorer
systems. \citet{finoguenov2000} used {\sl{ASCA}} data to show that SN~Ia
products dominate in the cluster centres and the \sncc\ products are more evenly
distributed (see the left panel of Fig.~\ref{radialfig}). 

However, the large and energy dependent point-spread function ($\sim$2\arcmin\
half-power radius) of {\sl{ASCA}} did not allow to investigate the spatial
abundance distributions in detail. \citet{degrandi2001} took advantage of the
better spatial resolution ($\sim$1\arcmin\ half-power radius) of
{\sl{BeppoSAX}} and they measured the radial Fe abundance profiles for a
sample of 17 rich nearby clusters of galaxies. They found that while the eight
non-cooling core clusters in their sample have flat Fe abundance profiles, the
Fe abundance is enhanced in the central regions of the cooling core clusters
(see the right panel of Fig.~\ref{radialfig}). \citet{degrandi2004} show that
the mass associated with the abundance excess found at the centre of cool core
clusters can be entirely produced by the brightest cluster galaxy (BCG). 

The bandpass of the {\sl{ROSAT}} PSPC was well suited for X-ray bright galaxies
and groups of galaxies with temperatures of $\sim$1~keV and its PSF allowed for
spatially resolved spectral analysis on a half-arcminute scale.
\citet{buote2000b} investigated the deprojected abundance gradients in 10
galaxies and groups. In 9 of the 10 systems they found an abundance gradient,
with the Fe abundance one to several times the Solar value within
$\sim$10~kpc, decreasing to $\sim$0.5 Solar at 50--100~kpc distance. 

The emerging picture from the abundance studies with \asca\ and \beppo\ data was
that of an early homogeneous enrichment by \sncc, which produce
$\alpha$-elements in the protocluster phase, and a subsequent more centrally
peaked enrichment by SN~Ia which have longer delay times and continue to explode
in the cD galaxy for a long time after the cluster is formed.  

Analysing and stacking all the data of clusters of galaxies observed with
{\sl{ASCA}} into temperature bins, \citet{baumgartner2005} found a trend
showing that the relative contribution of SN$_{\mathrm{CC}}$ is larger in higher
temperature clusters than in the clusters with lower temperature. Moreover, they
found that Si and S do not track each other as a function of temperature, and Ca
and Ar have much lower abundances than expected. They conclude that the
$\alpha$-elements do not behave homogeneously as a single group and the trends
indicate that different enrichment mechanisms are important in clusters of
different masses.

\section{The possibilities and limitations of elemental abundance determinations}
\label{possibilitiessec}

\subsection{Spectral modelling and the most common biases}
\label{modelingsec}

\begin{figure}   
\begin{center}
\hspace{-1.5cm}
\begin{minipage}{0.45\textwidth}
\includegraphics[width=5.5cm,clip=t,angle=0]{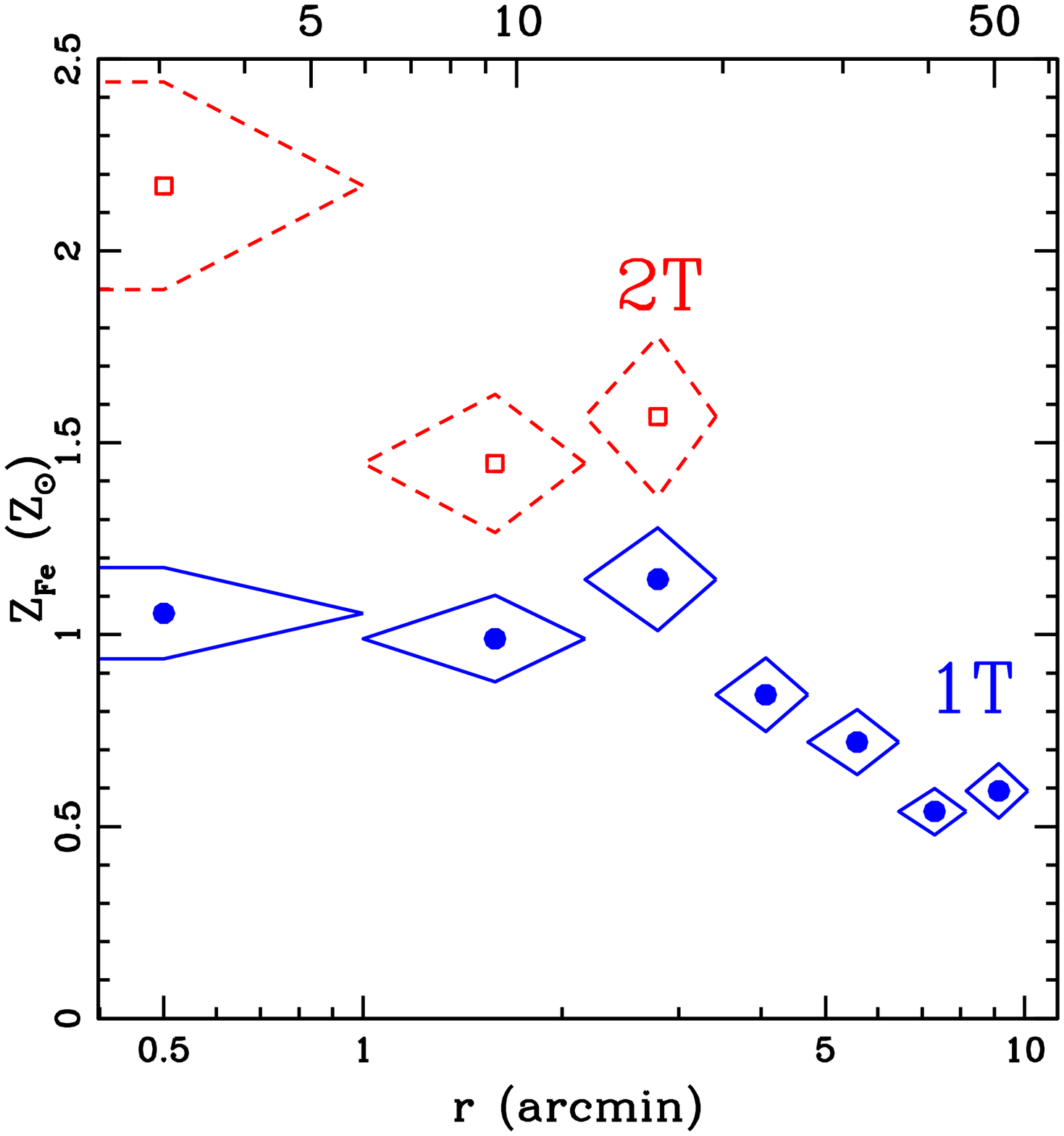}
\end{minipage}
\hspace{-0.1cm}
\begin{minipage}{0.45\textwidth}
\includegraphics[width=5.15cm,clip=t,angle=270]{fig5b.ps}
\end{minipage}

\caption{These plots illustrate the influence of the temperature model on the
best-fit abundances. \leftp Radial Fe abundance distribution in NGC~1399
modelled with single- and two-temperature models \citep[from][]{buote2002}.
\rightp Radial S abundance distribution in
the cluster 2A~0335+096 modelled with single, two, and multi-temperature models
\citep[triangles, squares, and circles, respectively; from][]{werner2006}. 
Abundances in both
figures are in the Solar units of \protect\citet{anders1989}.}

\label{biasesfig}
\end{center}  
\end{figure}

The correct modelling of the temperature structure is crucial for the
elemental-abundance determinations. Fitting the spectrum of a multi-temperature
plasma with a simple single-temperature model results in best-fitting
elemental-abundances that are too low. This effect is the most important in
galaxies, groups of galaxies and in the cooling cores of clusters. Many studies
have found significantly subsolar values for the Fe abundance in groups of
galaxies, with abundances generally less than the stellar values in these
systems, and less than those observed in galaxy clusters. To reconcile these
observations with the chemical enrichment models, it has been postulated that
galaxy groups accrete primordial gas after they have been expelling gas during
most of their evolution \citep{renzini1997}. \citet{buote1994} showed that if a
galaxy spectrum has intrinsically two temperature components, the
single-temperature model can give a significantly underestimated metallicity.
\citet{buote2000} demonstrated that in a spectrum characterised by two
temperatures, one below 1~keV and one above 1~keV, with an average of
$\sim$1~keV, the lower temperature component preferentially excites emission
lines  in the Fe-L complex below 1~keV (\ion{Fe}{xvii--xxi}), while the higher
temperature component excites the Fe-L lines from $\sim$1--1.4~keV
(\ion{Fe}{xxi--xxiv}). If these components contribute approximately equally to
the emission measure, the spectral shape of the Fe-L complex will be broader and
flatter than the narrower Fe-L complex produced by a single-temperature model
with the average temperature of the two components. In order to fit the flat
spectral shape at 1~keV, the single-temperature model will fit a lower Fe
abundance and will increase the thermal continuum. The strongest bias is at the
low temperatures, for higher temperature plasmas the bias is smaller as the Fe-K
line-emission becomes more important for the determination of the Fe abundance.
\citet{buote2000} also shows an excess in the Si/Fe and S/Fe ratios inferred for
two-temperature plasmas fit by single-temperature models. He notes that the lower
temperature component in the two-temperature plasma produces stronger Si and S
emission, and the single-temperature model underestimates the continuum at
higher energies. Therefore the Si/Fe and S/Fe ratios of multi-temperature ICM
inferred using single-temperature models will yield values in excess of the true
values. \citet{buote2000} shows that this effect is the most important for the
hotter clusters of galaxies. This bias may have played a role in the excess
Si/Fe ratios and trends derived using ASCA data
\citep{mushotzky1996,fukazawa1998,finoguenov2000,baumgartner2005}. 

Several authors \citep[e.g.][]{buote2003,werner2006,matsushita2007b}, using
XMM-Newton and Chandra data, showed that the best-fitting abundances in the
multi-phase cooling core regions of clusters and in groups increase when
two-temperature models or more complicated differential emission measure (DEM)
models are used. \citet{buote2003} show that the abundances of O, Mg, Si, and Fe
in the centre of NGC~5044 obtained using a two-temperature model are higher than
those obtained using a single-temperature model. They also show that the Si/Fe ratio in the centre is
significantly lower when a two-temperature model is used. Similar results were obtained for NGC~1399 \citep[see the left panel of Fig.~\ref{biasesfig}][]{buote2002}.
Analysing deep XMM-Newton data of the cluster 2A~0335+096, \citet{werner2006} showed that the
abundances of all fitted elements (Mg, Si, S, Ar, Ca, Fe, Ni) in their model
increased as they went from single- to two-temperature model, and to the
multi-temperature DEM model. However, not only their absolute values increased
but also the Si/Fe and the S/Fe ratios changed. \citet{matsushita2007b} show an
increase from single- to two- to multi-temperature models for O, Mg, Si, and Fe
in the core of the Centaurus cluster.  In the right panel of
Fig.~\ref{biasesfig} we show the radial S profile obtained for 2A~0335+096
\citep{werner2006} with single-, two-, and multi-temperature models. 

As discussed in \citet{buote2000b} and \citet{buote2003} in the lower mass
systems ($\sim 1$~keV) the best fit Fe abundance is sensitive to the lower
energy limit used in spectral fitting. If the continuum at the low energies is
not properly modelled, the equivalent widths of the lines in the Fe-L complex
(and the oxygen line emission) will be inaccurate and the abundance
determination will be incorrect. \citet{buote2003} show the best-fitting Fe
abundance in NGC~5044 determined with the lower energy limit at 0.7, 0.5, and
0.3~keV. They show that the results for each energy limit are significantly
different and they obtain the highest Fe abundance for the lowest energy limit,
which allows them to properly fit the continuum. 

Since the $\sim0.2$~keV component of the Galactic foreground emits \ion{O}{viii}
line emission, the uncertainties in the foreground/background model can cause
systematic uncertainty in determining the oxygen abundance in the ICM. The
temperature and flux of the $\sim0.2$~keV Galactic foreground component is
position dependent on the sky and  cannot be properly subtracted with blank
fields \citep[for detailed discussion see][]{deplaa2006}.

Systematic uncertainties in the abundance determination can be introduced if
bright point sources are not excluded from the spectral fits, especially in the
lower surface brightness regions of clusters. The contribution of point sources
raises the continuum level, systematically reducing the observed line equivalent
widths. If their contribution is not properly modeled, the best fitting thermal
model will have a higher temperature and lower elemental abundances. 

In general two plasma models are used for fitting cluster spectra: MEKAL
\citep{mewe1985,mewe1986,kaastra1992,liedahl1995,mewe1995} and APEC
\citep{smith2001}.  These codes have differences in their line libraries and in
the way of modelling the atomic physics. This allows to test the robustness of
the abundance determinations. Systematic effects on the best fit abundances due
to the adopted plasma model were investigated for example by \citet{buote2003},
\citet{sanders2006}, and \citet{deplaa2007}. They conclude that despite small
differences in the determined temperature structure, the derived abundances are
generally consistent within the errors.

Using simulated galaxy clusters \citet{kapferer2007b} show that the more
inhomogeneously the metals are distributed within the cluster, the less accurate
is the observed metallicity as a measure for the true metal mass. They show that
in general the true metal mass is underestimated by observations. If the
distribution of metals in the ICM is inhomogeneous, the true metal mass in the
central parts ($r<500$~kpc) of galaxy clusters can be up to three times higher
than the metal mass obtained by X-ray observations. The discrepancies are due to
averaging. As the metallicity and temperature are not constant throughout the
extraction area, the integration of thermal Bremsstrahlung and of line radiation
can lead to underestimated metal masses.

\subsection{Possibilities and limitations of current instruments}

{\sl{XMM-Newton}} with its large effective area and superb spectral resolution, {\sl{Chandra}} with its unprecedented spatial resolution, and {\sl{Suzaku}} with its low background and good spectral resolution at low energies largely enhanced our knowledge on chemical abundances in galaxies, groups, and clusters of galaxies. In this section we describe the possibilities and the systematic uncertainties of abundance studies with these instruments.

\subsubsection{XMM-Newton}
\label{xmmsec}

\begin{figure}   
\begin{center}
\includegraphics[height=0.8\textwidth,clip=t,angle=-90]{fig6a.ps}
\\
\vspace{5mm}
\includegraphics[height=0.8\textwidth,clip=t,angle=90]{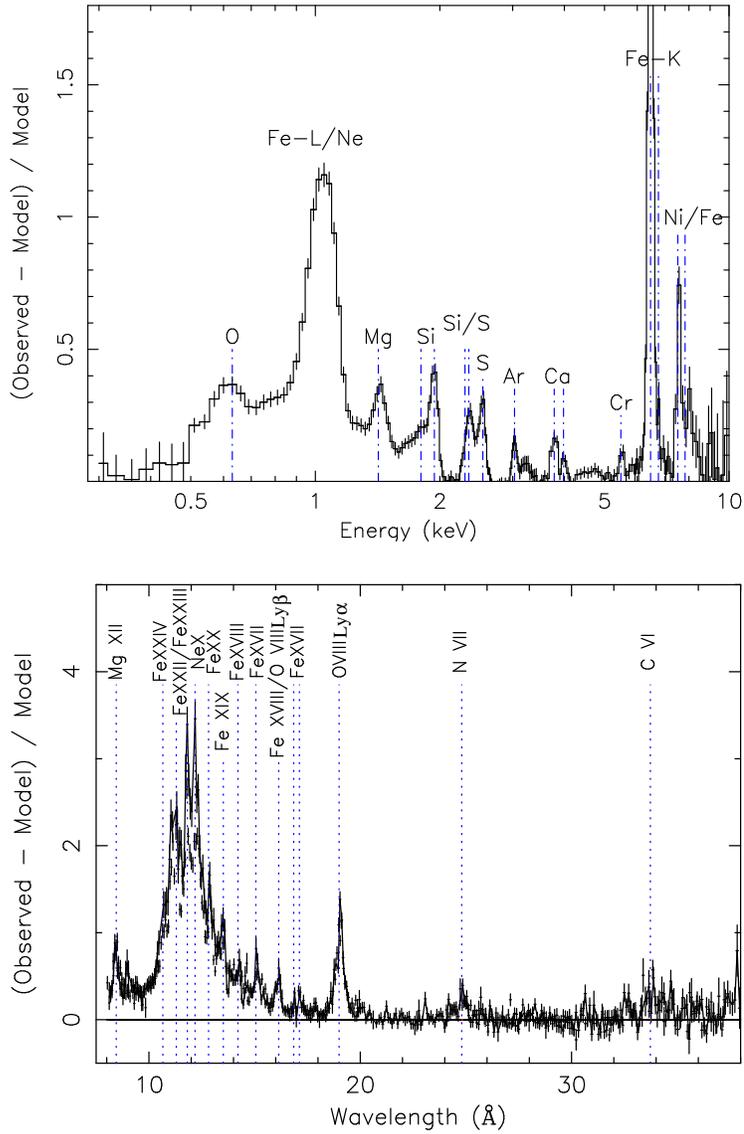}
\caption{{\sl{Top panel:}} The line spectrum of the cluster 2A~0335+096, as observed with {\sl{XMM-Newton}} EPIC \citep[from][]{werner2006}. 
{\sl{Bottom panel:}} Line spectrum of M~87, as observed with {\sl{XMM-Newton}} RGS \citep[from][]{werner2006b}. After the fitting, the line emission was set to zero in the multi-temperature plasma model indicated on the y-axis. While the CCD spectra are plotted as a function of the observed energy, the grating spectra are shown as a function of the observed wavelength.}
\label{XMMfig}
\end{center}  
\end{figure}

Due to its large effective area and good spectral resolution,
{\sl{XMM-Newton}} is currently the best suited instrument to study abundances
in the ICM. {\sl{XMM-Newton}} has three identical telescopes and two sets of
detectors on board. The first set of instruments is the European Photon Imaging
Camera (EPIC), consisting of two metal oxide semiconductor - MOS1 and MOS2 - CCD
arrays collecting 50~\% of light from two telescopes, and the EPIC-pn CCD array
collecting 100~\% of light from one telescope on board. The second set of
instruments are the Reflection Grating Spectrometers (RGS) that collect 50~\% of
light from two telescopes. Both sets of instruments play an important role in
the abundance studies.

The advantages of EPIC are its broad bandpass, and good spatial and spectral
resolutions, that allow us to detect the emission lines from O, Ne, Mg, Si, S,
Ar, Ca, Fe, and Ni (see the top panel of Fig.~\ref{XMMfig}). Furthermore, for
deep observations of bright clusters, EPIC allows us to derive the spatial
distribution of elements (see Sect.~\ref{spatialsec}). 

Unfortunately, the abundances determined in the cluster outskirts with low
surface brightness using \xmm\ remain uncertain due to the high background
level. The value of the best fit temperature in these regions is very sensitive
to the level of the subtracted background \citep[e.g.][]{deplaa2006} and as we
show above in Sect.~\ref{modelingsec}, the temperature modelling is very
important for the correct abundance determination. 

The O line emission in the spectrum is strong, but its abundance is difficult to
measure because it is sensitive to the assumed model for the Galactic foreground
(see Sect.~\ref{modelingsec}) and there are still remaining calibration
uncertainties at low energies. The 1s--2p Ne lines at 1.02~keV are in the middle
of the Fe-L complex (lying between 0.8 to 1.4~keV). The resolution of the EPIC
cameras is not sufficient to resolve the individual lines in the Fe-L complex,
which makes the Ne abundance determination by EPIC uncertain. However, the
abundances of O and Ne in the cores of cooling core clusters, observed with
\xmm\ with sufficient statistics, can be well determined using the high spectral
resolution of the RGS. 

The systematic differences in the effective-area between EPIC-MOS and EPIC-pn
are of the order of 5--10~\% in certain bands \citep{kirsch2006}. By using data
with high photon statistics and fitting spectra obtained by EPIC-MOS and EPIC-pn
separately, \citet{deplaa2007} investigated the magnitude of the systematic
errors on the best-fitting abundances of elements heavier than Ne due to the
uncertainties in the effective area calibration. They found that the elemental
abundances most influenced by the calibration uncertainties are Mg, Si, and Ni. 
\citet{deplaa2007} show that the systematic uncertainties on Mg are the largest,
since the effective-area of EPIC-pn has a dip with respect to the effective-area
of EPIC-MOS at the energies of the magnesium lines. This causes the Mg abundance
derived by EPIC-pn to be underestimated. For the abundances of Si and Ni they
estimate that the systematic uncertainties are 11~\% and 19~\%, respectively. 
Although EPIC-MOS and EPIC-pn show large systematic differences at the energy of
the Mg lines, the Mg abundances derived by EPIC-MOS and \chandra\ ACIS for the
Centaurus cluster are consistent \citep{matsushita2007b,sanders2006}, and the Mg
abundances derived by the EPIC-MOS and RGS using the deep observations of
2A~0335+096 also agree with each other. 

The S abundance determination with EPIC is robust~\citep[e.g.][]{deplaa2007}. Ar
and Ca abundances should also be accurately determined, however their signal is
usually weak, so the level of their observed equivalent widths may be somewhat
sensitive to the continuum level. 

The RGS has a high spectral resolution but only a limited spatial resolution in
the cross-dispersion direction. It allows for relatively accurate abundance
measurements of O, Ne, Mg, Fe, and in the case of deep observations of nearby
bright clusters and elliptical galaxies even the spectral lines of C and N are
detected (see the bottom panel of Fig.~\ref{XMMfig}). In the case of deep
observations of nearby bright clusters, RGS allows us to extract spectra from
typically five bins that are $\sim$1\arcmin\ wide in the cross-dispersion
direction and the photons in each bin are collected from $\sim$10\arcmin\ long
regions in the dispersion direction. This way RGS allows us to determine
projected abundance profiles.    

The spectral resolution of the RGS is, however, limited by the broadening of the
observed line profiles caused by the spatial extent of the source along the
dispersion direction. This makes RGS spectroscopy useful only for cooling core
clusters with strongly peaked surface brightness distribution. In order to
account for the line broadening in the spectral modelling, the model is
convolved with the surface brightness profile of the source along the dispersion
direction. Because the radial profile for a spectral line produced by an ion can
be different from the radial surface brightness profile in a broad band, in
practice the line profile is multiplied with a scale factor, which is a free
parameter in the spectral fit.  Different best fit line profiles for the Fe
lines and for the O line emission can suggest a different radial distribution
for these elements \citep{deplaa2006}.  

\subsubsection{Chandra}
\label{chandrasec}

Although the Advanced CCD Imaging Spectrometer (ACIS) on {\sl{Chandra}} has a
much smaller effective area than the EPIC on \xmm, which makes it less suitable
for abundance studies of the ICM, it is the instrument of choice for studies of
metal enrichment in galaxies, where a good spatial resolution is required.
{\sl{Chandra}} is able to resolve the substantial fraction of X-ray binaries
in galaxies as point sources, so their contribution can be excluded from the
spectral modelling of the inter-galactic medium (IGM).  The excellent spatial
resolution also makes it possible to investigate the spatial temperature
variations, which can bring a bias into the determination of abundances (see
Sect. \ref{modelingsec}). The spectral resolution of ACIS is sufficient to
reliably determine the abundances of a number of elements in the ICM and IGM.
With deep enough exposure times {\sl{Chandra}} ACIS can be used to study the
ICM abundances of O, Ne, Mg, Si, S, Ar, Ca, Fe, and Ni
\citep[e.g.][]{sanders2004,sanders2006}. 

\citet{humphrey2004} discuss the systematic uncertainties in the abundance
determinations due to the uncertainties in the absolute calibration of ACIS.
After correcting for the charge transfer inefficiency (CTI) using an algorithm
by \citet{townsley2002}, they found a change of $\Delta
Z_{\mathrm{Fe}}\simeq0.3$ in the best fit Fe abundance of NGC~1332. 

\citet{sanders2006} investigate the systematic uncertainties on the abundance
determinations with the S3 chip of {\sl{Chandra}} ACIS in the Centaurus
cluster. They conclude that for the best determined elements the
non-CTI-corrected {\sl{Chandra}} and {\sl{XMM-Newton}} results agree well.
The only element where the {\sl{Chandra}} and {\sl{XMM-Newton}} data do not
agree is oxygen. There has been a buildup of contamination on the ACIS
detectors, which influences the effective area calibration at the low energies
making the O abundance determination difficult. The iridium coating of the
{\sl{Chandra}} mirrors produces a steep drop in the effective area at around
2.05~keV, introducing a potential systematic uncertainty in the Si abundance
determination. However, the Si abundances determined by {\sl{Chandra}} and
{\sl{XMM-Newton}} agree well \citep{sanders2006}.

\subsubsection{Suzaku}
\label{suzakusec}

The advantage of the X-ray Imaging Spectrometer (XIS) detectors
\citep{koyama2007} on {\sl{Suzaku}} \citep{mitsuda2007} compared to EPIC on
{\sl{XMM-Newton}} is its better spectral resolution below 1~keV, and the low
background, providing a better sensitivity at the O line energy. {\sl{Suzaku}}
also allows for better measurements of the Mg lines in the low surface
brightness cluster outskirts, which is difficult with EPIC/MOS due to the
presence of a strong instrumental Al line close to the expected energy of the
observed Mg lines. 

The relatively broad point spread function of XIS with a half-power diameter of
$\sim$2\arcmin\ \citep{serlemitsos2007} does not allow detailed investigations
of the spatial distribution of elements. 

The effective area of the XIS detectors below 1~keV is affected by carbon and
oxygen contamination of the optical blocking filter. The contamination is time
and position dependent. Its column density and spatial distribution are still
not well known. \citet{matsushita2007} estimate that the uncertainty in the
thickness and chemical composition of the contaminating column results in a
$\sim$20~\% systematic error on the O abundance.

\section{Spatial distribution of elements}
\label{spatialsec}

\subsection{Radial abundance profiles}
\label{radialsec}

\begin{figure}    
\begin{center}
\begin{minipage}{0.45\textwidth}
\includegraphics[width=5.3cm,clip=t,angle=-90]{fig7a.ps}
\end{minipage}
\begin{minipage}{0.45\textwidth}
\hspace{1.75cm}
\includegraphics[width=4.2cm,clip=t,angle=0]{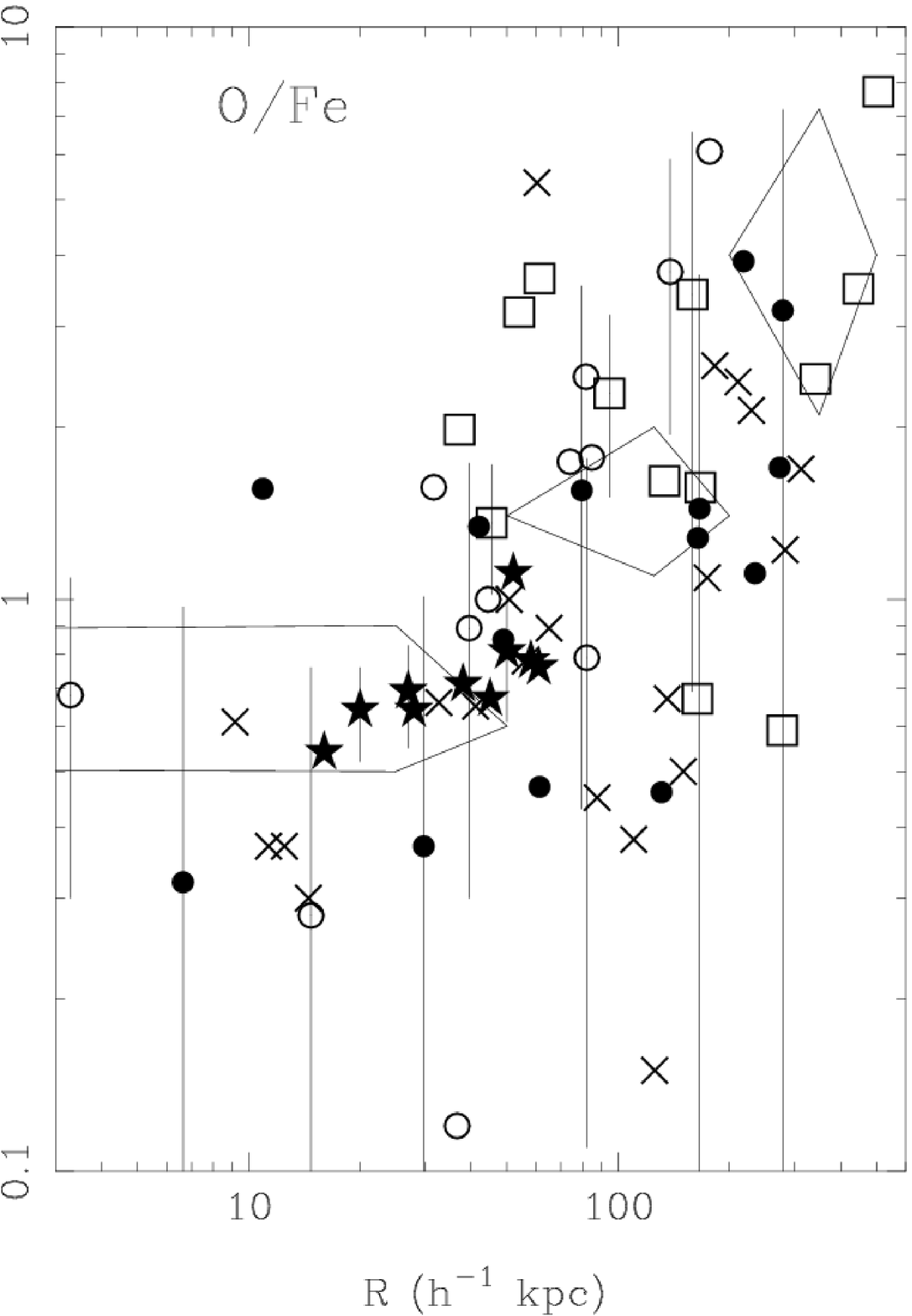}
\end{minipage}
\caption{\leftp Radial distribution of Fe and O in M~87 determined using \xmm\ RGS \citep[from][]{werner2006b}. The abundances are shown relative to the proto-solar values given by \citet{lodders2003}. \rightp Radial distribution of the O/Fe ratio determined for a sample of clusters using \xmm\ EPIC \citep[from][]{tamura2004}. The abundance ratios are shown relative to the Solar values given by \citet{anders1989}.}
\label{abunproffig}
\end{center}  
\end{figure}

Observations with \xmm\ and \chandra\ confirmed the centrally peaked metallicity
distribution in the cooling core clusters and the flat distribution of metals in
the non-cooling core clusters \citep{vikhlinin2005,pratt2007} previously seen
with {\sl{BeppoSAX}}. \xmm\ and \chandra\ also confirmed the strong
contribution of SN~Ia to the enrichment of cluster cores found by \asca. The
observations are consistent with the picture of an early contribution by \sncc\
and later enrichment of cluster cores by SN~Ia. One of the first abundance
studies with \xmm, performed on the cooling core cluster Abell~496, showed a
radially flat distribution of O, Ne, and Mg and centrally peaked distribution of
Ar, Ca, Fe, and Ni \citep{tamura2001b}. The abundance peak in the cluster core
is consistent with the idea that the excess metals were produced by SNe~Ia in
the cD galaxy. However, the strongly increasing Si/Fe ratio toward the outskirts
seen with \asca\ was not confirmed with \xmm. 

Analysing \xmm\ EPIC data of M~87, \citet{bohringer2001} and
\citet{finoguenov2002} found rather low O/Fe ratio in the cluster core, that
increased toward the outer regions. Analysing the same dataset,
\citet{gastaldello2002} questioned these results, showing that the O/Fe ratio in
the inner 9\arcmin\ is constant. As shown in the left panel of
Fig.~\ref{abunproffig}, using \xmm\ RGS data, \citet{werner2006b} found
a centrally peaked Fe abundance gradient and a flat O distribution in M~87
confirming the results of \citet{bohringer2001}, \citet{finoguenov2002}, and
\citet{matsushita2003} (the O abundance profiles found by
\citet{gastaldello2002} were different because of calibration problems at low
energies in the early days of the mission). The RGS data reveal an O/Fe ratio of
0.6 in the cluster core and a ratio of 0.9 at a distance of 3\arcmin\ (14~kpc)
from the centre. However, the \xmm\ EPIC data show that the Si abundance has a
similar gradient in M~87 as the Fe abundance. A similar abundance pattern with a
rather flat radial distribution of O, and centrally peaked Si, and Fe abundances
is observed in the cluster sample analysed by \citet[][see the right panel of
Fig.~\ref{abunproffig}]{tamura2004}, in the Perseus cluster \citep{sanders2004},
in Abell~85 \citep{durret2005}, in S\'ersic~159-03 \citep{deplaa2006}, in
2A~0335+096 \citep{werner2006}, in the Centaurus cluster
\citep{matsushita2007b,sanders2006}, and in the Fornax cluster
\citep{matsushita2007b}. The abundances of Si and Fe have similar gradients also
in the group of galaxies NGC~5044 \citep{buote2003}, with a constant radial
distribution of Si/Fe$=0.83$. The radial distribution of O in this group is
consistent with being flat. {\sl{Suzaku}} observations of the clusters A~1060
\citep{sato2007a} and AWM~7 \citep{sato2007b} also showed rather low O abundance
of around 0.5 Solar within about 5\arcmin\ from the core without a clear
abundance decline in the outer regions.

The radial metallicity gradients in cooling flow clusters often display an
inversion in the centre \citep[e.g.][]{sanders2002}. This apparent metallicity
drop in the very central region is often the result of an oversimplified model
in the spectral analysis \citep[see Sect.~\ref{possibilitiessec}
and][]{molendi2001}. However, in some cases the metallicity drop in the cluster
core is robust. In the Perseus cluster the abundances peak at the radius of
40~kpc and they decrease at smaller radii. This inversion does not disappear
when extra temperature components and power-law models are included, or when
projection effects are taken into account \citep{sanders2007}. 

\citet{bohringer2004} found that long enrichment times ($\gtrsim$5~Gyr) are
necessary to produce the observed central abundance peaks. Since the classical
cooling flows, or a strongly convective intra-cluster medium, or a complete metal
mixing by cluster mergers would destroy the observed abundance gradients, they
conclude that cooling cores in clusters are preserved over very long times. 
\citet{bohringer2004} and \citet{matsushita2007} show that the
iron-mass-to-light ratio in the central region of the Centaurus cluster (the
mass of Fe relative to the B-band luminosity of the stellar population of the cD
galaxy) is much higher than that of M~87. This indicates that the accumulation
time scale of the SN~Ia products was much higher in Centaurus than in M~87.  In
agreement with the conclusions of \citet{degrandi2004} they also conclude that
the innermost part of the ICM is dominated by gas originating mainly from the
stellar mass loss of the cD galaxy. 

If the metals originate from the stellar population of the cD galaxy, then in
the absence of mixing the abundance profiles would follow the light profiles.
The observed metal profiles are much less steep than the light profiles, which
suggests that the injected metals are mixed and they diffuse to larger radii.
\citet{rebusco2005,rebusco2006} model the diffusion of metals by stochastic gas
motions after they were ejected by the brightest galaxy and compare their
results with the peaked Fe abundance profiles of 8 groups and clusters. They
suggest that the AGN/ICM interaction makes an important contribution to the gas
motion in the cluster cores. 

In order to determine the metallicities of the stellar populations of galaxies
and star clusters it has become a common practice to use the feature made up of
the Mg\,b line and the Mg H band around 5200~\AA, which is known as the Mg$_2$
index. The Mg$_2$ index appears to be a relatively good metallicity indicator.
In M~87, Centaurus, and the Fornax clusters the O and Mg abundances in the
centre are consistent with the optical metallicity of the cD galaxy derived from
the Mg$_{2}$ index \citep{matsushita2003,matsushita2007,matsushita2007b}. The
Mg$_{2}$ index mainly depends on the Mg abundance, but also on the total
metallicity where the O contribution matters most, and depends weakly on the age
population of stars.  The agreement between the O and Mg abundances in the ICM
with the stellar metallicities supports the view that the ICM in the cluster
centres is dominated by the gas ejected from the cD galaxies.  However, the
Mg/Fe ratio in the stellar populations of elliptical galaxies is significantly
higher than that in the hot gas \citep{humphrey2006}. The observed difference
shows that since the time when the current stellar populations of these
ellipticals formed, a significant number of SNe~Ia enriched the gas in addition
to the stellar mass loss.   

Observations indicate that at large radii clusters have a flat abundance
distribution of about 0.2 Solar. By using {\sl{Suzaku}} data, \citet{fujita2007}
found that the gas is enriched to a metallicity of $\sim$0.2 Solar out to the
virial radii of clusters A~399 and A~401. These clusters are at the initial
stage of a merger and their virial radii partially overlap. The authors
interpret their finding as a result of early enrichment by galactic superwinds
in the proto-cluster stage. However, the extraction region used by
\citet{fujita2007} is large and their measured Fe abundance might be
significantly biased toward the value at about half of the virial radii of the
clusters.

\subsection{2D distribution of metals}
\label{2Dsec}

Deep observations of bright clusters of galaxies with \xmm\ and \chandra\ allow
to map the 2D distribution of metals in the ICM
\citep{sanders2004,durret2005,o'sullivan2005,sauvageot2005,finoguenov2006,werner2006,sanders2006,bagchi2006}.
In general the distribution of metals in clusters is not spherically symmetric,
and in several cases it shows edges in the abundance distribution or several
maxima and complex metal patterns. The 2D distribution of metals is a good
diagnostic tool, which can be used together with the maps of thermodynamic
properties to investigate the complex structure of cluster cores and the merging
history of clusters \citep[e.g.][]{kapferer2006}.

\begin{figure}    
\begin{center}
\includegraphics[width=8.5cm,clip=t,angle=0.]{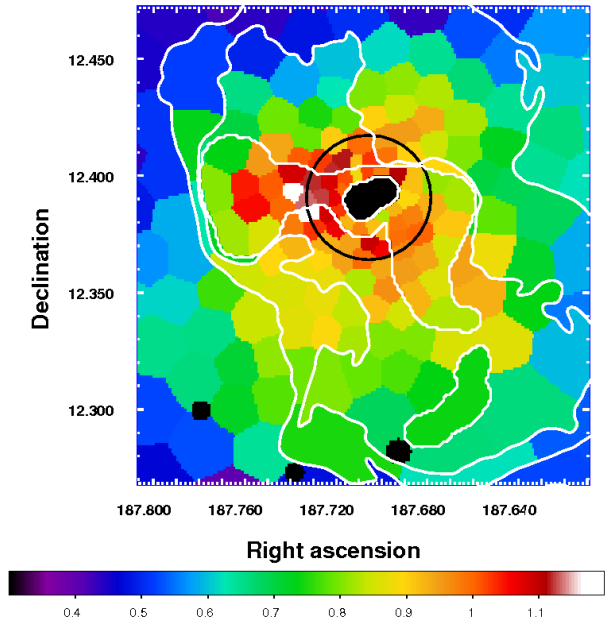}
\caption{Map of the Fe abundance in M~87. The colour bar indicates the Fe
abundance with respect to the proto-solar value of \protect\citet{lodders2003}, with red
showing higher and blue lower abundances. Contours of the 90 cm radio emission
\citep{owen2000} are overplotted. The black circle indicates the half light
radius of M~87. Beyond the expected radial gradient, one clearly sees the
enhanced Fe abundance in the radio arms, especially within the Eastern arm. From
\citet{simionescu2007b}.}
\label{M87metals}
\end{center}  
\end{figure}

M~87 provides us with the unique opportunity to study in detail the role of the
AGN feedback in transporting and distributing the metals into the ICM.  The high
X-ray surface brightness and small distance of M~87 allow a detailed study of
the 2D distribution of several elements (O, Si, S, and Fe). By analysing deep
(120~ks) XMM-Newton data, \citet{simionescu2007b} found that the spatial
distribution of metals shows a clear enhancement along the radio lobes, where
the rising radio emitting plasma ejected by the AGN uplifts cooler gas from the
core of the cluster (see Fig.~\ref{M87metals}). Furthermore, they showed that
the metallicity of the cool uplifted (${\mathrm k}T<1.5$~keV) gas is $\sim$2.2
times Solar and the abundance ratios in and outside the multi-phase
region associated with the uplifted gas are very similar. They estimate that the
mass of the cool gas is $\approx$$5\times10^8$~$M_{\odot}$ and it probably
originates from the stellar winds enriched with SN~Ia products. Approximately
$2\times10^8$~yr are required to produce the observed mass of cool gas,
indicating that the uplift of cool gas by AGN radio bubbles is a rare event.

\section{Element ratios and their reconstruction with supernova models}
\label{ratiosec}

The chemical elements that we see in the ICM are the integral yield of all the
different supernova types that have left their specific abundance patterns in
the gas prior and during cluster evolution. Since the launch of {\sl{ASCA}},
which for the first time allowed the determination of abundances of elements
other than Fe in the ICM (see Sect.~\ref{historysec}), several groups tried to
use the ICM abundance patterns to constrain the contribution of different kinds
of supernovae to the ICM enrichment and to put constraints on the theoretical
supernova models.  

In the following subsection we review these efforts. The still somewhat
ambiguous results of these efforts are summarised in section 6.2

\subsection{Results on reconstructed supernova models}

\citet{mushotzky1996} investigated the elemental abundances in four clusters,
removing the central regions, and concluded that the abundance ratios are
consistent with the contribution of \sncc.   \citet{dupke2000} found that most
of the Fe ($\sim60-70$~\%) in the cores of their investigated clusters comes
from SN~Ia, and they used these central regions dominated by SN~Ia ejecta to
discriminate between competing theoretical models for SN~Ia. They showed that
the observed high Ni/Fe ratio of $\sim$4 is more consistent with the W7
convective deflagration SN~Ia model than with the delayed detonation models.
Stacking all the cluster data in the {\sl{ASCA}} archive into temperature
bins, \citet{baumgartner2005} showed that the abundances of the most well
determined elements Fe, Si, and S do not give a consistent solution for the
fraction of material produced by SN~Ia and SN$_{\mathrm{CC}}$ at any given
cluster temperature. They concluded that the pattern of elemental abundances
requires an additional source of metals other than SN~Ia and \sncc, which could
be the early generation of Population~III stars. 

The {\sl{ASCA}} estimates were largely based on the Si/Fe ratio, which has the
disadvantage of Si being contributed by both SN~Ia and SN$_{\mathrm{CC}}$. The
much improved sensitivity of {\sl{XMM-Newton}} to measure the spectral lines
of more elements at a higher statistical significance offers better
opportunities to constrain the supernova models. 

\citet{finoguenov2002} used \xmm\ EPIC to accurately measure the abundances of
O, Ne, Mg, Si, S, Ar, Ca, Fe, and Ni in the hot gas in two radial bins of M~87.
They found that the Si-group elements (Si, S, Ar, Ca) and Fe have a stronger
peak in the central region than O, Ne, and Mg. The SN~Ia enrichment in the
centre is characterised by a Solar ratio of Si-group elements to Fe. The
inferred SN~Ia enrichment in the outer region, associated with the ICM of the
Virgo cluster, has a lower ratio of Si-group elements to Fe by 0.2 dex, which is
more characteristic for the ICM of other clusters \citep{finoguenov2000}.
\citet{finoguenov2002} also find a Ni/Fe ratio of $\sim$1.5 in the central
region, while they point out that this ratio for other clusters determined by
{\sl{ASCA}} is $\sim$3. They conclude that these abundance patterns confirm
the diversity of SN~Ia and to explain the SN~Ia metal enrichment in clusters
both deflagration and delayed detonation scenarios are required.  The high SN~Ia
yield of Si-group elements in the centre may imply that the Si burning in SN~Ia
is incomplete. This favours the delayed-detonation models with lower density of
deflagration-detonation transition, lower C/O ratio, and lower central ignition
density. On the other hand, the abundance patterns in the outer region of M~87
are more characteristic of deflagration supernova models. 

A bimodal distribution is observed in the magnitude decline of the SN~Ia events.
While a population of SN~Ia with a slow decline is more common in spiral and
irregular galaxies with recent star formation, a fainter and more rapidly
decaying population of SN~Ia is more common in early-type galaxies
\citep{hamuy1996,ivanov2000}. The SN~Ia with incomplete Si burning enriching the
central part of M~87 could be associated with optically fainter SNe~Ia observed
in early-type galaxies.  \citet{mannucci2006} found that the present data on
SN~Ia rates in different populations of parent galaxies indicate that SN~Ia have
a bimodal delay time distribution. They conclude that about 50~\% of SN~Ia
explode soon after the formation of the progenitor binary system, in a time of
the order of $10^{8}$ years, while the delay time distribution of the remaining
50~\% of SN~Ia can be described by an exponential function with a decay time of
$\sim$3~Gyr. The supernovae with short delay times might be associated with the
supernovae with the slower decline, while the supernovae with long delay times
might be associated with the rapidly decaying SNe~Ia. If this association is
true, then the observed Si/Fe of the accumulated SN~Ia ejecta will be a function
of the age of the system. 

\citet{matsushita2007b} show that SN~Ia ejecta have a higher Fe/Si ratio in
regions with larger iron-mass-to-light ratio (IMLR). Higher IMLR corresponds to
longer accumulation time scale. The correlation between the inferred Fe/Si ratio
of SN~Ia ejecta and the IMLR supports the suggestion by \citet{finoguenov2002}
and \citet{matsushita2003} that the average Fe/Si ratio produced by SN~Ia
depends on the age of the system. 

\citet{boehringer2005b} compiled the available observational results of the
O/Si/Fe from \xmm, \chandra, and \asca\ to show that the observed abundance
patterns in clusters are more consistent with the WDD models, which provide
larger Si/Fe ratios than the W7 model. \citet{buote2003} and
\citet{humphrey2006} also discuss the higher $\alpha$-element enrichment
observed in groups and in elliptical galaxies. They conclude that there is an
increasing evidence that the O is over-predicted by the theoretical
SN$_{\mathrm{CC}}$ yields, or there is a source of $\alpha$-element enrichment
in addition to that provided by SN$_{\mathrm{CC}}$ and SN~Ia. 

\begin{figure}    
\begin{center}
\begin{minipage}{0.45\textwidth}
\vspace{0.5cm}
\includegraphics[width=5.9cm,clip=t,angle=0]{fig9a.ps}
\end{minipage}
\hspace{0.25cm}
\begin{minipage}{0.45\textwidth}
\includegraphics[width=5.9cm,clip=t,angle=0]{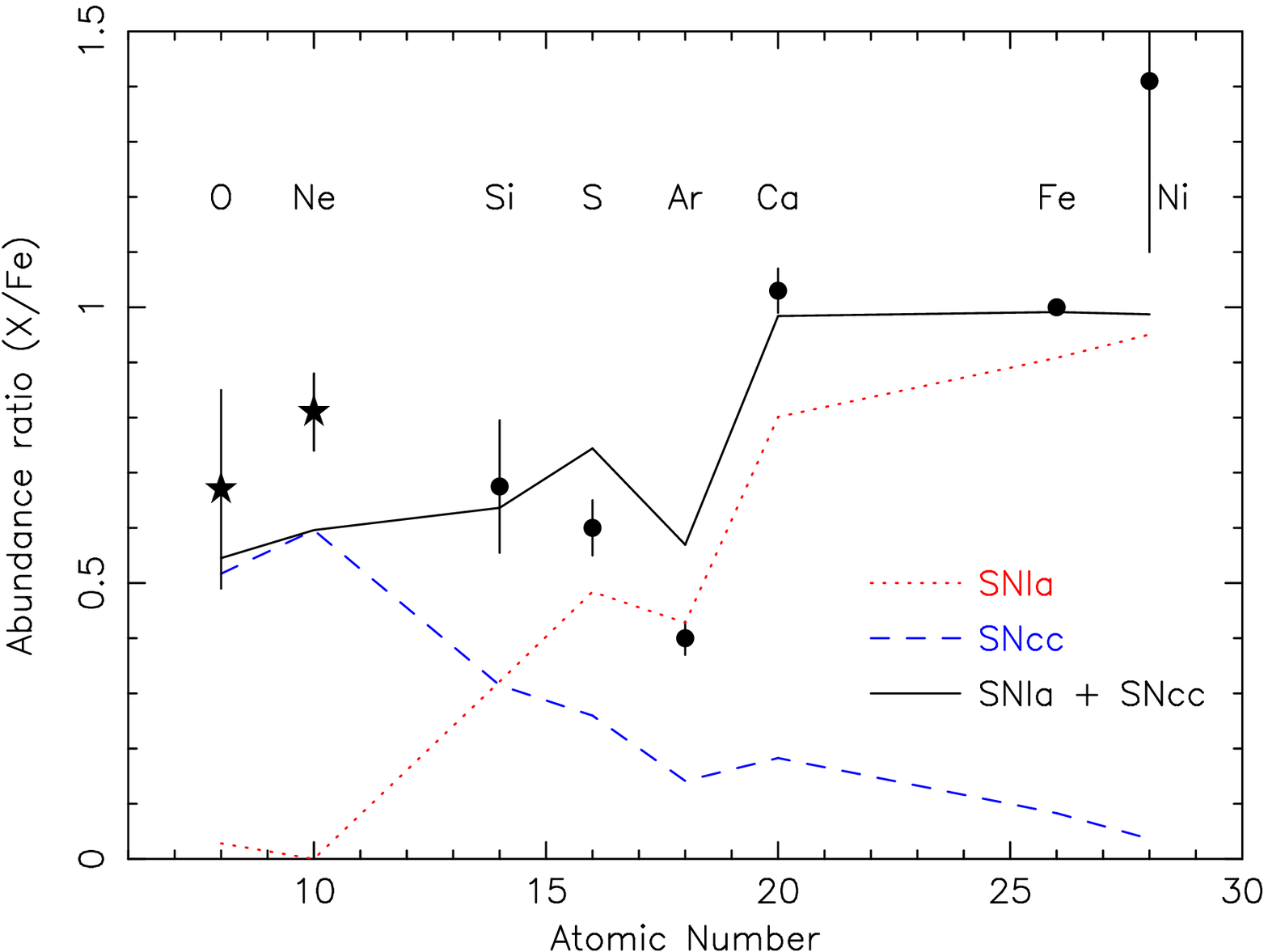}
\end{minipage}

\caption{\leftp\ Comparison of the [Fe/Si] ratio of the ICM in the Centaurus
cluster and in M~87, plotted against [Fe/O], from \citet{matsushita2007b}. 
Abundances are here on the scale of \protect\citet{feldman1992}. The
open square shows the average value for Galactic metal poor stars
\citep{clementini1999} and the asterisk shows the abundance ratios for \sncc\
calculated by \citet{nomoto1997}. The solid line and dashed line show the
abundance pattern synthesised by SN~Ia with Fe/Si=1.1 (best fit relation for the
core of M~87) and Fe/Si=2.6 \citep[the ratio produced by the W7
model][]{nomoto1984}, respectively. 
\rightp\ Reconstruction of the abundance
patterns (at the proto-solar \protect\citealt{lodders2003} scale) observed in the sample of 22~clusters of \citet{deplaa2007} with
theoretical supernova yields. The abundance values indicated by filled circles
are obtained with \xmm\ EPIC. The abundances of O and Ne are the average values
obtained for S\'ersic 159-03 and 2A~0335+096 obtained with RGS. The SN~Ia yields
are for a delayed detonation model calibrated on the Tycho supernova remnant by
\citet{badenes2006}. The \sncc\ yields are for progenitors with Solar
metallicities by \citet{nomoto2006} integrated over the Salpeter IMF. From
\citet{deplaa2007}.}
\label{deplaafig}
\end{center}  
\end{figure}

Using deep \xmm\ observations of the clusters 2A~0335+096 and S\'ersic~159-03,
\citet{werner2006} and \citet{deplaa2006} determined the abundances of 9
elements and fitted them with nucleosynthesis models for supernovae. They found
that $\sim$30~\% of all supernovae enriching the ICM were SN~Ia and $\sim$70~\%
were SN$_{\mathrm{CC}}$. They also found that the Ca abundance in these clusters
is higher than that expected from the models. Later, \citet{deplaa2007} used a
sample of 22 clusters observed with {\sl{XMM-Newton}} to determine the
abundances of Si, S, Ar, Ca, Fe, and Ni in the ICM within the radius of
$0.2R_{500}$ and they compared the best-fitting abundances with theoretically
predicted yields of different SN~Ia and SN$_{\mathrm{CC}}$ models (see
Fig.~\ref{deplaafig}).  Because the SN$_{\mathrm{CC}}$ do not have a significant
impact on the Ar/Ca ratio, \citet{deplaa2007} used the Ar and Ca abundances in
clusters as a criterion for determining the quality of SN~Ia models. They found
that the Ar and Ca abundances in the ICM are inconsistent with currently favored
SN~Ia models. However, they showed that their best fit Ca/Fe and Ar/Ca ratios
are consistent with the empirically modified delayed-detonation SN~Ia model,
which is calibrated on the Tycho supernova remnant \citep{badenes2006}. Using
this model, \citet{deplaa2007} obtained a number ratio of
$N_{\mathrm{SN\,Ia}}/N_{\mathrm{SN\,Ia+SN_{\mathrm{CC}}}}=0.44\pm0.10$, which
suggests that binary systems in the appropriate mass range are very efficient
($\sim 5-16$~\%) in eventually forming SN~Ia explosions. 

Based on {\sl{Suzaku}} observations of 2 clusters (A1060 and AWM~7) and 2 groups
of galaxies (HCG~62 and NGC~507), \citet{sato2007c} obtained abundance patterns
for O, Mg, Si, S, and Fe in the region out to $0.3r_{180}$. Contrary to most of
the previously mentioned results, they found a better fit to the observed
abundance patterns using the W7  SN~Ia model rather than the WDD models.
Assuming the W7 model for SN Ia, the yields from \citet{nomoto2006} and the
Salpeter IMF for SN$_{\rm CC}$, the ratio of occurrence numbers of SN$_{\rm cc}$
to SN Ia is $\sim$3.5. It corresponds to the number ratio of
$N_{\mathrm{SN\,Ia}}/N_{\mathrm{SN\,Ia+SN_{\mathrm{CC}}}}=0.22$, which is smaller
than that obtained by \citet{deplaa2007}. The number ratios were determined only
considering the metals in the ICM and metals locked up in the stars were not
considered. 

\subsection{Summary of the efforts of using ICM abundances to constrain
supernova models}

The abundance patterns obtained during many deep observations of
clusters of galaxies with XMM-Newton clearly favor the delayed-detonation SN~Ia
models \citep[see][]{boehringer2005b} and the observed abundance ratios of Ar/Ca
are relatively well reproduced using empirically modified delayed-detonation
models calibrated on the Tycho supernova remnant \citep{deplaa2007}. However,
several observations still favor the W7 SN~Ia model \citep{sato2007c} and in
most of the cases the abundances are not determined well enough to really
discriminate between the different models. As \citet{finoguenov2002} suggest it
might even be possible that both the deflagration and delayed-detonation types
of SN~Ia play a role in the chemical enrichment of clusters of galaxies.

{\sl{XMM-Newton}} RGS observations of the hot gas in clusters of galaxies and
in elliptical galaxies reveal Ne/O ratios which are higher than Solar
\citep{xu2002,peterson2003,werner2006b,werner2006}. SN~Ia produce very little O
and Ne, and their ratio can be used to put constraints on the progenitors of
SN$_{\mathrm{CC}}$. From the SN$_{\mathrm{CC}}$ models discussed in
Sect.~\ref{supernovasec} only the models with pre-enriched progenitors predict
higher than Solar Ne/O ratio.

While it is well known that elements from O up to the Fe-group are primarily
produced in supernovae, the main sources of C and N are still being debated (see
Sect.~\ref{supernovasec}). Deep observations of nearby X-ray bright elliptical
galaxies and bright clusters with the RGS on \xmm\ allow to measure the C and N
abundances in the hot gas. By combining the RGS data obtained in two deep \xmm\
observations of M~87, \citet{werner2006b} determined relatively accurate C and N
abundance values. The small O/Fe ratio and large C/Fe and N/Fe ratios found in
M~87 suggest that the main sources of C and N are not the massive stars that
also produce large quantities of O, but the low- and intermediate-mass
asymptotic giant branch stars.

\section{Abundances as function of cluster mass and redshift}
\label{masssec}

\citet{arnaud1992} studied the correlations between the gas mass, Fe mass, and
optical luminosity in clusters of galaxies. They found that the gas mass in the
ICM is highly correlated and the Fe mass is directly proportional to the
inegrated optical luminosity of elliptical and lenticular galaxies. They found
no correlation with the integrated optical luminosity of spiral galaxies. The
ratio of the gas mass in the ICM to the stellar mass increases with the cluster
richness.

\citet{renzini1997} showed that the IMLR is very similar in all clusters of
galaxies with ICM temperature above $\sim$1 keV.  However, below $\sim$1~keV the
IMLR seems to drop by almost 3 orders of magnitude. This drop at low
temperatures is due to the combination of lower gas mass and lower measured Fe
abundance in cool systems. However, it was shown that the low Fe abundances
measured in many groups were artifacts of wrong temperature modelling (see
later). 

\citet{renzini1997,renzini2004} showed that clusters hotter than $\sim$2.5~keV
have similar abundances, with very small dispersion. Below 2.5~keV, the best fit
metallicities show a large range of values. The previously reported very low
abundance values for low temperature groups and elliptical galaxies turn out to
be much higher when two- or multi-temperature fitting is applied \citep[see
Sect.~\ref{modelingsec} and][]{buote1998,buote2000,buote2003,humphrey2006}.
However, the outer regions of galaxy groups still show very low metallicities.
An XMM-Newton observation of the group NGC~5044 indicates an iron abundance in
the hot gas of $Z_{\mathrm{Fe}}\sim 0.15$ Solar in the region between 0.2 and
0.4 of the virial radius \citep{buote2004b}. This result shows that the total
iron mass to optical light ratio in NGC~5044 is about 3 times lower than that
observed in rich clusters. 

\begin{figure}
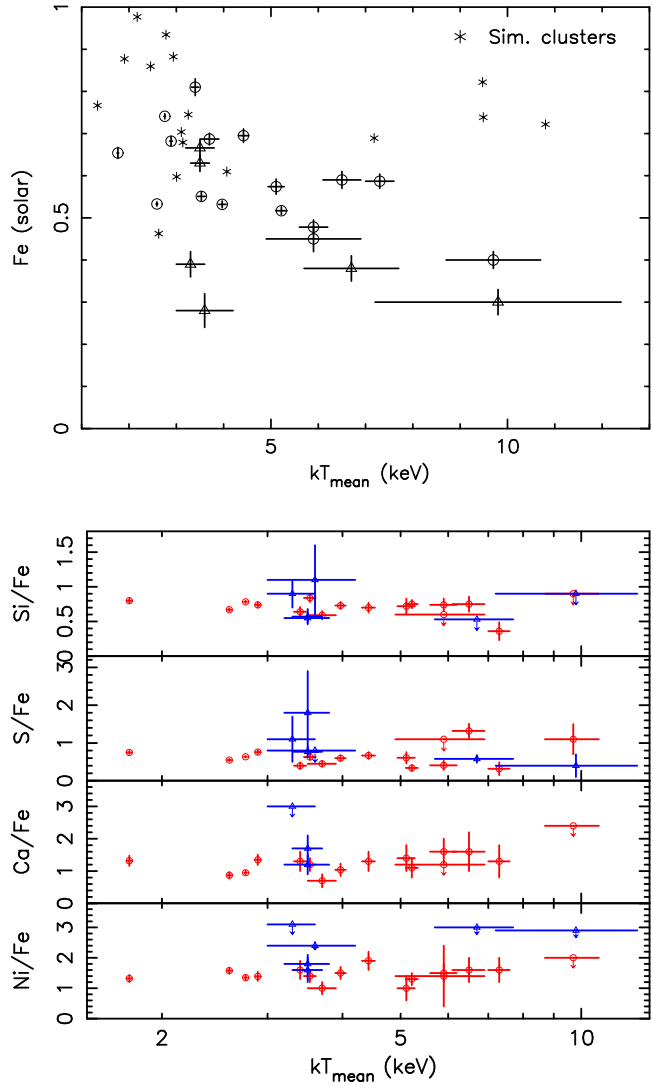
    
\begin{center}
\includegraphics[height=0.7\textwidth,clip=t,angle=-90]{fig10a.ps}
\\
\vspace{5mm}
\includegraphics[height=0.7\textwidth,clip=t,angle=-90]{fig10b.ps}
\caption{{\sl Top panel:} Best fit Fe abundance as a function of ICM temperature
determined within 0.2$R_{500}$ for a sample of clusters investigated by
\citet{deplaa2007}. We compare the observed Fe abundances with values from
simulations (\citealt{borgani2008} - Chapter 18, this volume) 
for a Salpeter IMF, extracted within the same physical radius. 
{\sl Bottom panel:} Abundance ratios of Si, S, Ca, and Ni with respect to Fe 
as a function of the cluster temperature, compiled from \citet{deplaa2007}. 
The abundance ratios are consistent with being constant as a 
function of cluster mass. 
Cooling core clusters are shown as red circles and non-cooling core 
clusters as blue triangles. Abundances in both panels are plotted with
respect to the 
proto-solar abundance values of \protect\citet{lodders2003}.}
\label{jelleratios}
\end{center}  
\end{figure}

We compiled the abundance values measured within 0.2$R_{500}$ for the sample of
22 clusters analysed by \citet{deplaa2007}.  In the top panel of
Fig.~\ref{jelleratios} we show the best fit Fe abundance as a function of ICM
temperature. Our data confirm earlier results showing that while in cooler
clusters, in the temperature range of $2$--$4$~keV, $Z_{\mathrm{Fe}}$ shows a
range of values between $0.2$--$0.9$ Solar, in hot massive clusters
(k$T\gtrsim5$~keV) the Fe abundance is on average lower and equal to
$Z_{\mathrm{Fe}}\sim0.3$ Solar. We compare the observed trend with values from
simulations extracted within 0.2$R_{500}$ as the {\sl{XMM-Newton}} data
(\citealt{borgani2008} - Chapter 18, this volume). The data are simulated for
the Salpeter IMF.  The data are more consistent with the simulated values at the
lower temperatures. At higher temperatures the simulated abundances are higher
than the observed values. The observed trend might be linked to a decrease of
the star formation efficiency with increasing cluster mass \citep{lin2003}.

In the bottom panel of Fig.~\ref{jelleratios} we show the abundance ratios of Si,
S, Ca, and Ni with respect to Fe as a function of the cluster temperature,
compiled from \citet{deplaa2007}. The abundance ratios are consistent with being
constant as a function of temperature, which is related to the cluster mass. 
The intrinsic spread in abundance ratios between clusters is smaller than 30~\%.
This finding contradicts the {\sl{ASCA}} results \citep{baumgartner2005}, which
show an increase of the Si/Fe ratio from 0.7 to 3 Solar in the $2$--$8$~keV
temperature range. This result indicates that the ratios of SN~Ia and
SN$_{\mathrm{CC}}$ contributing to the enrichment of the ICM are similar for
clusters of all masses. 

\begin{figure}    
\begin{center}
\begin{minipage}{0.45\textwidth}
\hspace{-1.25cm}
\includegraphics[width=5.2cm,clip=t,angle=0]{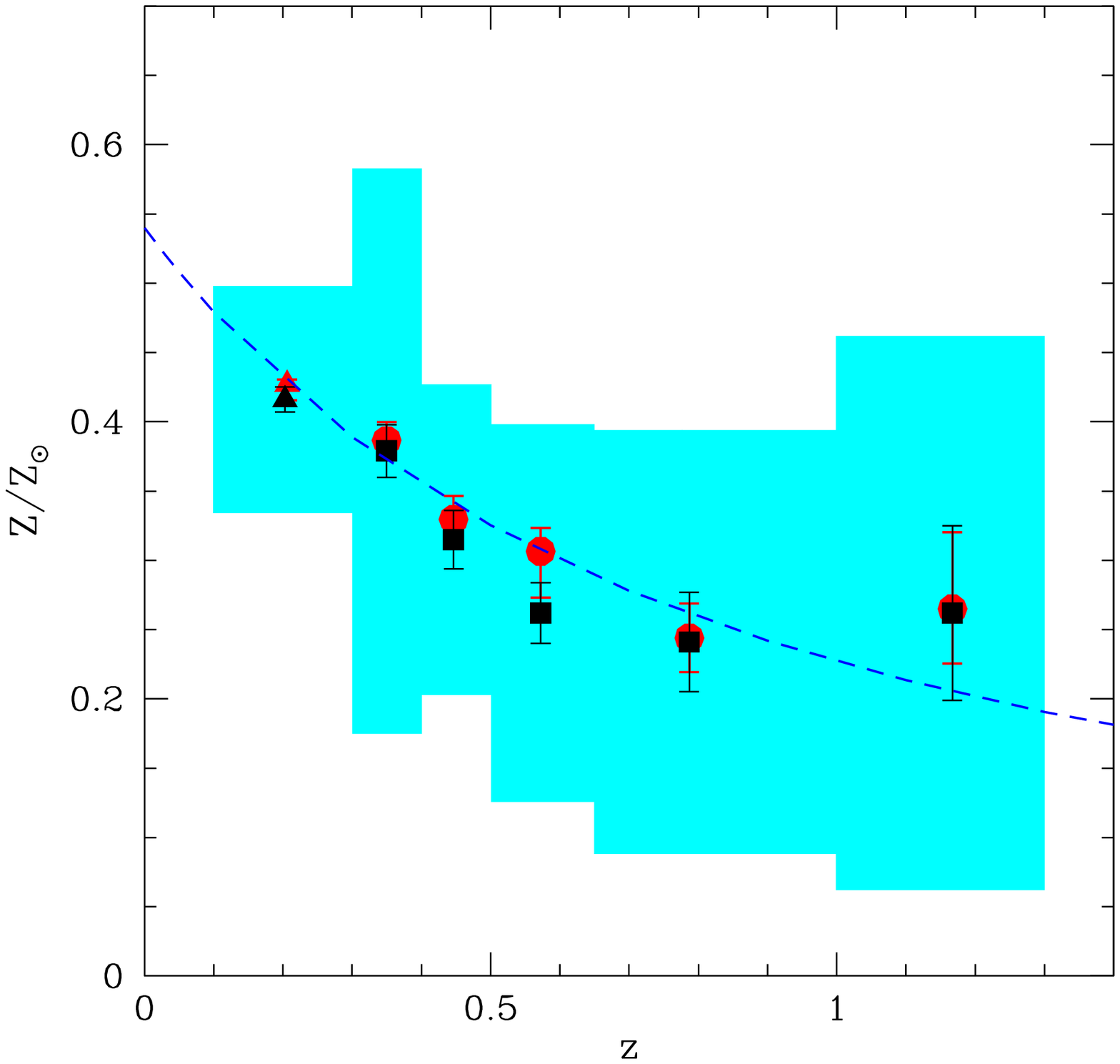}
\end{minipage}
\hspace{-1.75cm}
\begin{minipage}{0.45\textwidth}
\includegraphics[width=4.8cm,clip=t,angle=270]{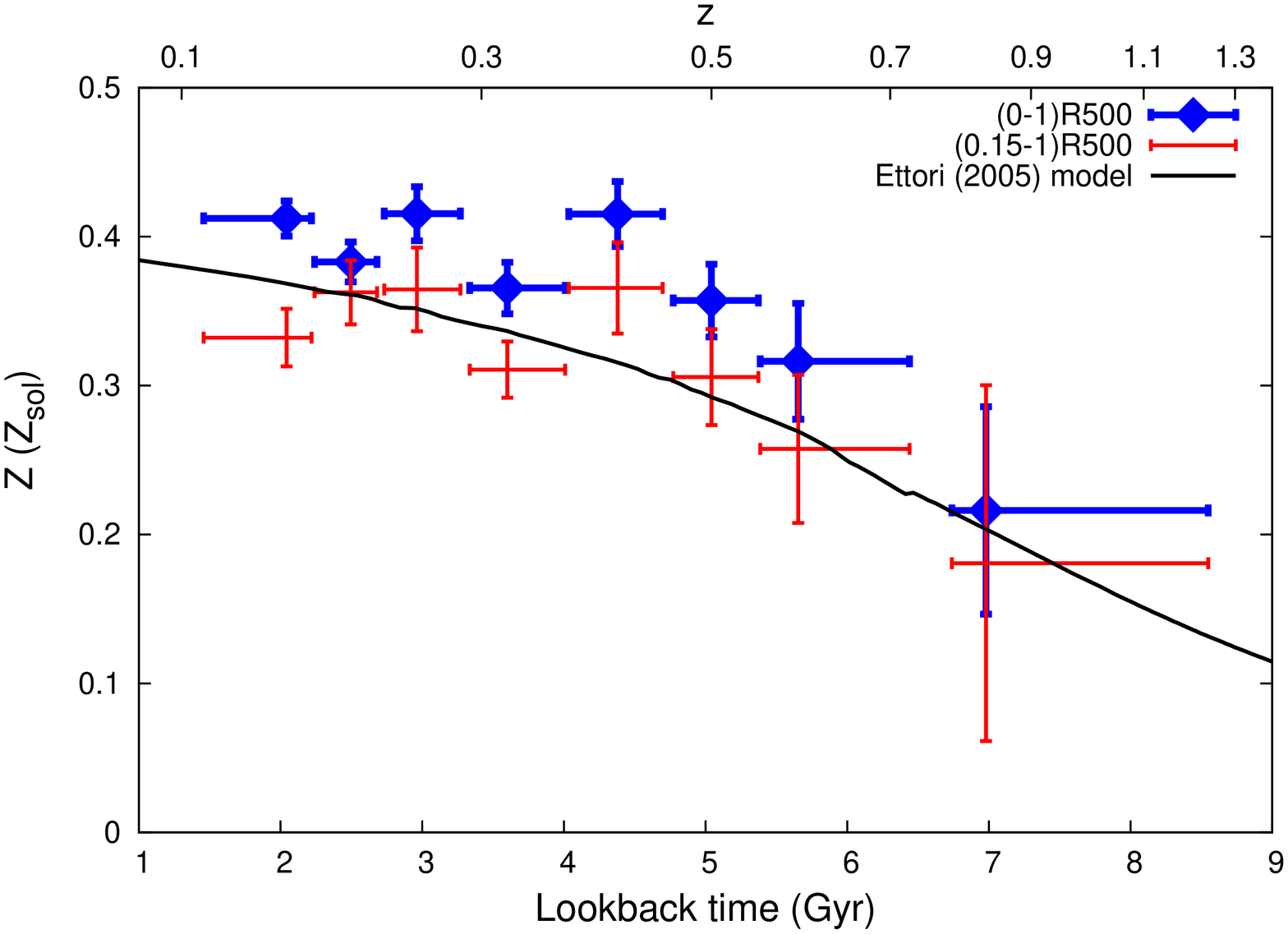}
\end{minipage}
\caption{\leftp\ Mean Fe abundance from combined fits within six redshift bins
(red circles) and the weighted average of single source observations in the same
bins (black square) from \citet{balestra2007}. The shaded areas show the $rms$
dispersion. The dashed line shows the best power-law fit over the redshift bins.
\rightp\  Mean ICM metal abundances in eight redshift bins within $R_{500}$ with
and without the central $0.15R_{500}$ excluded \citep[from][]{maughan2007}. The
solid line shows the supernova enrichment model of \citet{ettori2005}. Both
panels use the abundance scale of \protect\citet{anders1989}.}
\label{Fezevolution}
\end{center}  
\end{figure}

Investigating a sample of 56 clusters at $z\gtrsim0.3$, \citet{balestra2007}
confirmed the trend of the Fe abundance with cluster temperature also in the
higher redshift clusters. They show that the abundance measured within
(0.15--0.3)$R_{\mathrm{vir}}$ in clusters below 5~keV is, on average, a factor
of $\sim$2 larger than in hot clusters, following a relation
$Z(T)\simeq0.88T^{-0.47}$. 

\citet{balestra2007} also found an evolution of the Fe content in the ICM with
redshift (see the left panel of Fig.~\ref{Fezevolution}). They conclude that the
average Fe content of the ICM at the present epoch is a factor of $\sim$2 larger
than at $z\simeq1.2$.  While at the redshift $z\gtrsim0.5$ they observe an
average ICM abundance of $Z_{\mathrm{Fe}}\simeq0.25$ Z$_{\odot}$, in the
redshift range of $z\simeq0.3$--$0.5$ they measure
$Z_{\mathrm{Fe}}\simeq0.4$~Z$_{\odot}$. They parametrise the decrease of
metallicity with redshift with a power-law $\sim(1+z)^{-1.25}$.  The evolution
in the Fe abundance with redshift was confirmed by \citet{maughan2007}, who
investigated a sample of 116 clusters of galaxies at $0.1<z<1.3$ in the
{\sl{Chandra}} archive (see the right panel of Fig.~\ref{Fezevolution}). They
found a significant evolution with the abundance dropping by 50~\% between
$z\sim0.1$ and $z\sim1$. They verified that the evolution is still present, but
less significant, when the cluster cores are excluded from the abundance
measurement (they use apertures of 0--1 R$_{500}$ and 0.15--1 R$_{500}$),
indicating that the evolution is not solely due to disappearance of relaxed,
cool core clusters at $z\gtrsim0.5$. 

Simulations by \citet{kapferer2007a} show that between redshifts of $z=1$ and
$z=0$ a galaxy cluster accretes fresh gas, with the gas mass increasing during
this time period by a factor of $\approx$3. Which means that in order to get an
increase of the metallicity during this time period the freshly accreted gas
must be pre-enriched or the input of metals into the ICM in this time interval
must be larger than expected.

\section{Future of ICM abundance studies}
\label{future}

\begin{figure}    
\begin{center}
\includegraphics[width=0.7\textwidth,clip=t,angle=0]{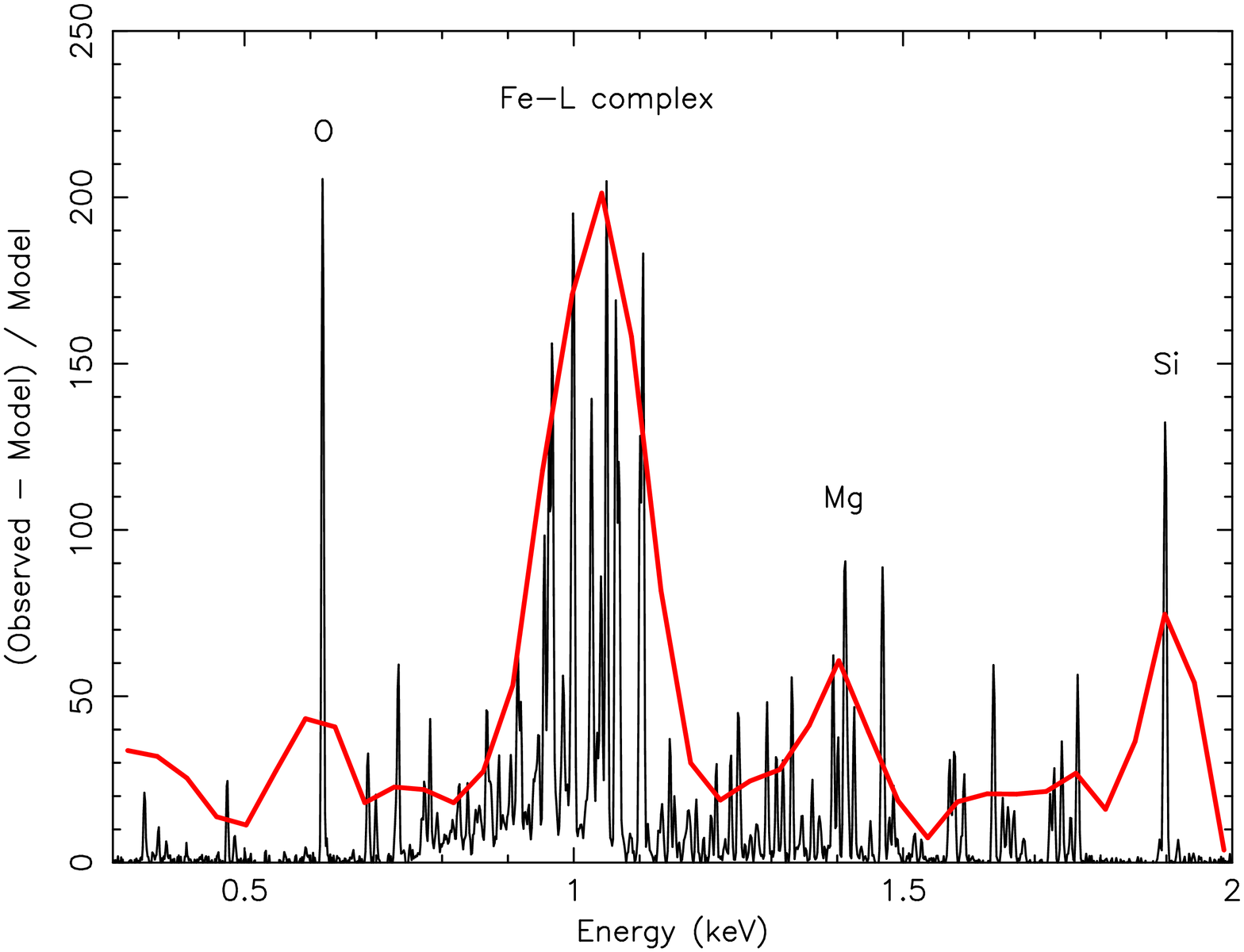}
\includegraphics[width=0.7\textwidth,clip=t,angle=0]{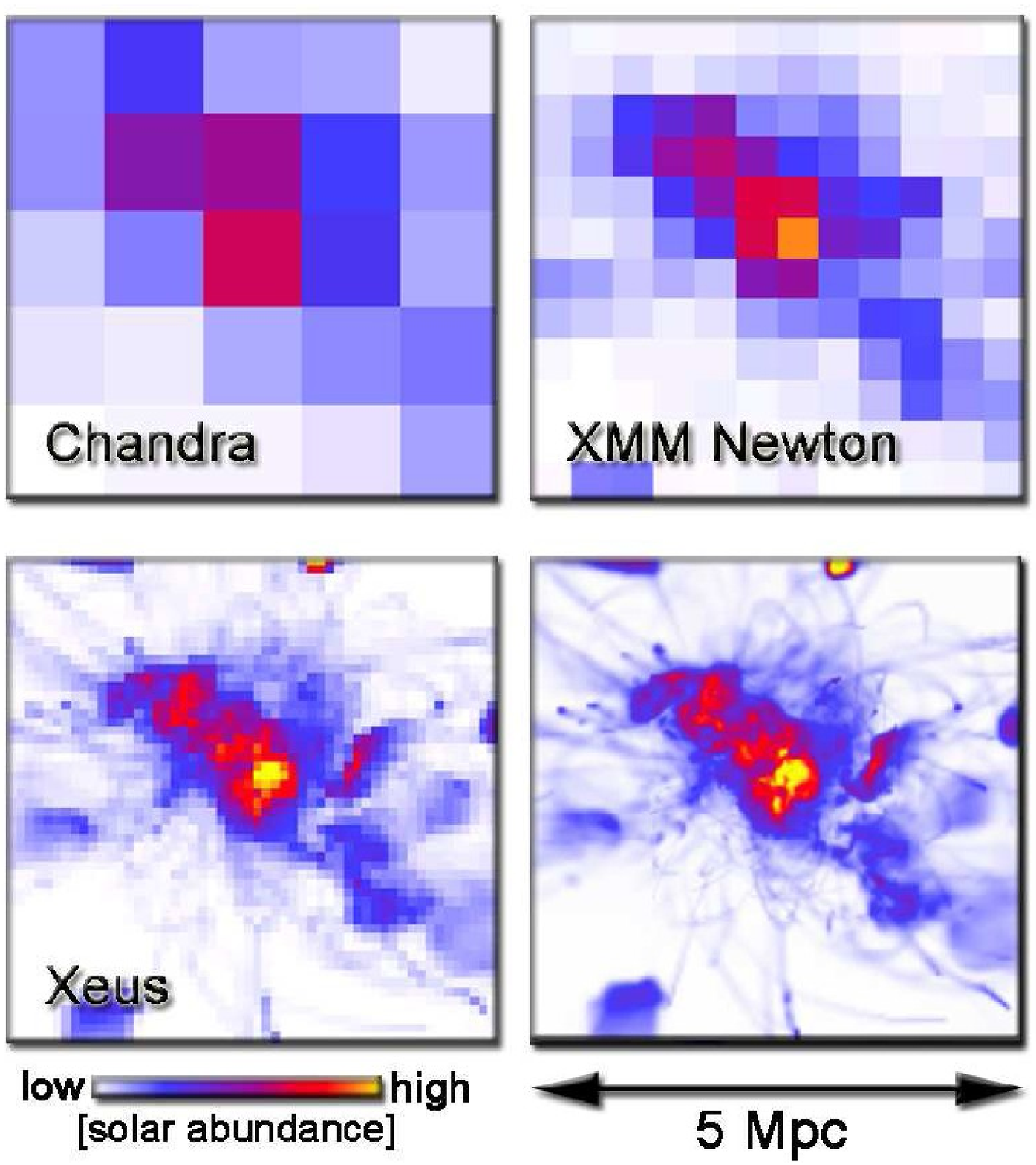}
\caption{{\it Top panel:} Simulated line spectrum of the inner 30\arcmin
\arcmin\ of the cluster of galaxies S\'ersic~159-03 for a 100~ks observation
with the TES array planned for XEUS. The continuum model is subtracted. For
comparison the EPIC spectrum of the same source is overplotted as the thick
line. From \citet{deplaa2007b}. {\it Bottom panel:} Simulated X-ray metallicity
maps as seen with \chandra, \xmm, and \xeus, respectively. The plot in the
bottom right corner shows the original simulation results. From
\citet{kapferer2006}. }
\label{xeusfig}
\end{center}  
\end{figure}

A major improvement in X-ray spectroscopy will be the employment of
micro-calorimeters, and especially of the so-called Transition-Edge Sensors
(TES) on future satellites (for more information on future instrumentation see
\citealt{paerels2008} - Chaper 19, this volume). Compared to the current
state-of-the-art CCD technology, this new type of detectors promises a factor of
$>20$ improvement in spectral resolution ($\lesssim 2-5$~eV, depending on the
optimalisation of the detector). To reach this high spectral resolution, the TES
detectors use the transition edge between normal conductivity and
superconductivity that is present in a number of materials. At temperatures of
$\sim$0.1~K, the conductivity of the absorbing material is very sensitive to the
temperature. When an X-ray photon hits the detector, its temperature rises and
the resistance of the material increases. By measuring the increase of the
resistance, one can calculate with high accuracy the energy of the X-ray photon.
The main effort today is to build an array of TES detectors, with imaging
capabilities. Next major proposed missions with TES arrays on board are, amongst
others, the {\sl{X-ray Evolving Universe Explorer}} ({\sl{XEUS}}), {\sl{Explorer
of Diffuse emission and Gamma-ray burst Explosions}} ({\sl{EDGE}}), and
{\sl{Constellation-X}}. 

The combination of the large effective area ($\sim$5~m$^2$) and high spectral
resolution ($\lesssim 2-5$~eV) of {\sl{XEUS}} will open up new possibilities in
cluster abundance studies, with enormous improvement in the abundance
determinations. In the left panel of Fig.~\ref{xeusfig} we show a simulated
spectrum of the cluster S\'ersic~159-03 \citep[from][]{deplaa2007b}, for the
planned TES detector that will be part of the Narrow-Field Imager (NFI) on
{\sl{XEUS}}. For comparison also the model at the resolution of \xmm\ EPIC is
shown. With {\sl{XEUS}} the abundances will be measured with an accuracy of
$\sim$$10^{-3}$ Solar, which is more than an order of magnitude improvement
compared to EPIC. This accuracy might be sufficient to detect the contribution
of the Population-III stars to the enrichment of the ICM. Also spectral lines
from more elements will be resolved, which will enable better tests for
supernova models. 

The large collecting area of {\sl{XEUS}} in combination with the CCD-type
detectors of the Wide-Field Imager will be excellent to observe the spatial
distribution of metals, which will help us to constrain the enrichment history
of clusters. Simulated metallicity maps by \citet{kapferer2006} for \chandra,
\xmm, and {\sl{XEUS}} are shown in the right panel of Fig.~\ref{xeusfig}. 

The proposed {\sl{EDGE}} satellite would also open up new possibilities in the
chemical abundance studies.  The combination of the planned long exposure times
(of the order of few Ms), large field of view (1.4 degrees) and good angular
resolution (Half Power Diameter of 15\arcmin \arcmin) of its CCD type Wide-Field
Imager, and the high spectral resolution (3 eV at 0.6 keV) of the Wide-field
Spectrometer promises new possibilities in the studies of metal abundances in
clusters. {\sl{EDGE}} would not only increase the number of detected spectral
lines and improve the accuracy of the abundance measurements, but it would also
enable us to accurately measure the chemical abundances and the spatial
distribution of metals in the faint outskirts of clusters out to their virial
radius, which is beyond the possibilities of current instruments. This would
provide us with insights of crucial importance for the understanding of the
chemical enrichment processes, nucleosynthesis, and of the metal budget of the
Universe.

\begin{acknowledgements}
The authors thank ISSI (Bern) for the support of the team ``Non-virialized X-ray
components in clusters of galaxies''. We would like to thank Hans B\"{o}hringer
and Alexis Finoguenov for reading the manuscript and providing comments.  NW
acknowledges support by the Marie Curie EARA Early Stage Training visiting
fellowship. The Netherlands Institute for Space Research is supported
financially by NWO, the Netherlands Organization for Scientific Research. We
thank Joop Schaye for his help with the core collapse supernova yields. RW
acknowledges support from Marie Curie Excellence Grant  MEXT-CT-2004-014112.
\end{acknowledgements}

\bibliographystyle{aa}
\bibliography{16_werner}

\end{document}